\documentclass[pr,twocolumn,amsmath,amssymb,floatfix]{revtex4}
\usepackage{color} 
\usepackage{graphicx}
\usepackage{dcolumn}
\usepackage{bm}
\usepackage{longtable}
\usepackage{textcomp}
\usepackage{epstopdf}
\usepackage{xcolor}

\begin{document}

\setlength{\parskip}{0 pt}

\bibliographystyle{unsrt}

\title{\centering Diffraction without Waves: \\ Emergence of the Quantum Substructure of Light}

\author{Joachim St\"ohr}

\address{SLAC National Accelerator Laboratory and
 Department of Photon Science,\\  Stanford, California 94035, USA }

\begin{abstract}
Today, the nature of light is accounted for by one of the jewels of physics, quantum electrodynamics (QED), the fundamental theory of light and matter. Yet owing to its infinite complexity, scientists still debate how its central concept, the photon, can be reconciled with the perceived existence of light waves, emerging 200 years ago in the wake of Young's double slit diffraction experiment. Ever since, the phenomenon of diffraction has been viewed to embody the wave nature of light, leading to the schizophrenic wave-particle duality. The latter does not exist in QED which is \emph{photon} based without the existence of waves.  Here we introduce the new paradigm that diffraction images directly reflect the fundamental quantum states of light. This is revealed by analysis of the evolution of modern versions of Young's experiment performed  with differently modified laser light and photon-based detection. In conventional quantum mechanics, corresponding to first order QED, the fundamental photon nature of light remains hidden since different quantum states produce only two basic types of diffraction patterns that may also be explained by coherent and incoherent wave superposition. The true photon based substructure of light is shown to clearly emerge through characteristic diffraction images in second order QED. The degeneracy of the  first order images is lifted, the wave-particle equivalence breaks down, and the patterns directly reveal the true quantum substructure of light. This allows the replacement of the conventional concept of wave coherence by a precise order-dependent degree of coherence that quantifies the interference and diffraction behavior of all quantum states of light.
\end{abstract}

\maketitle

\section{Introduction}

The description of the nature and behavior of light has arguably been one of the most studied problems in physics. Even today different experimental results are still explained  by two fundamentally different concepts based on the classical wave and quantum mechanical photon descriptions. The origin of this wave-particle ambiguity becomes most apparent in the description of light \emph{diffraction}, since the conventional double-slit diffraction pattern can be equally explained by the wave theory and by quantum mechanics.

The simplest quantum formulation of diffraction is due to Feynman \cite{feynman3} who explained Young's double slit experiment by use of his space-time probability amplitude formulation of quantum mechanics (QM) \cite{Feynman:48}. His treatment of diffraction may be viewed more generally as a demonstration of the inherent wave-particle duality underlying various formulations of quantum mechanics \cite{styer:2002}, first expressed through de Broglie's hypothesis that all matter has wave properties \cite{deBroglie:25}. Relative to other formulation of QM, Feynman's formulation is particularly appealing since  single photon probability amplitudes closely resemble  classical wave fields \cite{liu:2010,stohr-AOP}. Rather than resolving the wave-particle conflict, Feynman's treatment effectively consolidated it.

All formulations of diffraction within the confines of conventional QM are limited, however, by its well-known linearity \cite{Pang-Feng}. In particular, QM may be viewed as  a \emph{first order} perturbation within the complete theory of light and matter, quantum electrodynamics (QED), which extends to infinite order. Of the three different formulations of QED by Tomonaga \cite{tomonaga:46}, Schwinger \cite{schwinger:48} and Feynman \cite{Feynman:49}, it is again Feynman's formulation that is most appealing and of practical utility, as pointed out by Dyson \cite{dyson:1949} in showing their equivalence in 1949.

In particular, the concept of space-time probability amplitudes of single independent photons or electrons in QM may be extended to higher perturbative orders  through the construction of probability amplitudes of an increasing number of particles. Remarkably, in first order the elementary building blocks of QED, photons and electrons, are described by the \emph{same} probability amplitudes. This is the deeper reason why photons and electrons give the same conventional diffraction patterns \cite{feynman3,feynman-QED}. It is only in \emph{second order} that the description of multi-particles states becomes different for bosons (photons) and fermions (electrons), reflected for fermions by increasingly complex Feynman diagrams \cite{dyson:1949,feynman-QED}.

In Feynman's formulation, the probability amplitudes of an increasing number of photons, corresponding to increasing orders in QED, are constructed by addition and multiplication of those of individual photons \cite{liu:2010,stohr-AOP}.  The formulation for different cases is augmented by rules regarding the addition versus multiplication of single-particle amplitudes \cite{feynman-QED,stohr-AOP}. These rules become increasing complex for more than two photons and it is advantageous to use a different method of constructing multi-photon quantum states. This formalism, pioneered by Glauber \cite{glauber:63,glauber:63b,glauber:65}, underlies the modern formulation of quantum optics \cite{scully-zubairy,loudon}.

Glauber introduced the description of light in terms of orders of coherence $O\!=\!1,2,3...\infty$ \cite{glauber:65} which are equivalent to the orders of perturbation in Feynman's formulation of QED \cite{feynman-QED,stohr-AOP}. In Glauber's formulation, the construction of multi-photon probability amplitudes is facilitated by the use of an increasing number of photon birth and destruction (detection) operators. When for a given order the corresponding operators are applied to different quantum states one gets different expectation values. This leads to the link of quantum states and diffraction patterns, the central theme of the present article.

The existence of  quantum states of light containing specific number of photons $N$, was first experimentally verified by experiments in the late 1980s where individual photons ($N\!=\!1$) \cite{grangier:86} or photon pairs ($N\!=\!2$) \cite{Mandel:87} were sent through a lossless beam splitter and their emergence from different output ports was examined by coincidence detection, as reviewed in \cite{Ou:2007,Shih:2011,Ou:2017,stohr-AOP}. In the process it became clear that Dirac's famous statement that photons do not interfere with each other \cite{Dirac-book} holds only in first order QED. This was more directly revealed by diffraction experiments in the late 1990s where the conventional illumination of Young's double slits was replaced by use of entangled photon pairs \cite{pittman:95,Shih:2011}, produced by parametric down conversion \cite{klyshko:67,harris:67,magde-mahr:67}.

In this article, we introduce the general new paradigm that within QED, diffraction images are direct signatures of different quantum states of light. This becomes apparent when the results of modern versions of Young's double slit experiment, performed  by illumination with differently modified laser light and photon-based detection, are compared to the patterns predicted by the formulation of diffraction within QED.

This direct link has remained hidden in the past because the treatment of diffraction by conventional quantum mechanics results in an accidental degeneracy of diffraction patterns for different quantum states. Diffraction has therefore continued to be explained by the \emph{ad hoc} concepts of  \emph{``coherent''} and \emph{``incoherent''} superposition of waves. Here we show that the degeneracy of patterns for different quantum states in first order is lifted upon extension of QED to second order. This evolution is shown to be particularly important since the wave-particle equivalence breaks down and the true photon-based nature of light emerges in the diffraction patterns.

As a consequence, the wave theory of diffraction can in principle be abandoned altogether today, and the framework of statistical optics \cite{Born-Wolf,Goodman-SO} may be replaced by the more fundamental quantum formulation of light \cite{glauber:65,scully-zubairy}. In particular, the broad and difficult concept of ``partial coherence'' in wave optics can now be succinctly defined through the degrees of coherence of specific quantum states in different orders of QED.

\section{Generation of Different States of Light}
\label{SS:experiments}

\begin{figure}[!h]
\centering
\includegraphics[width=0.8\columnwidth]{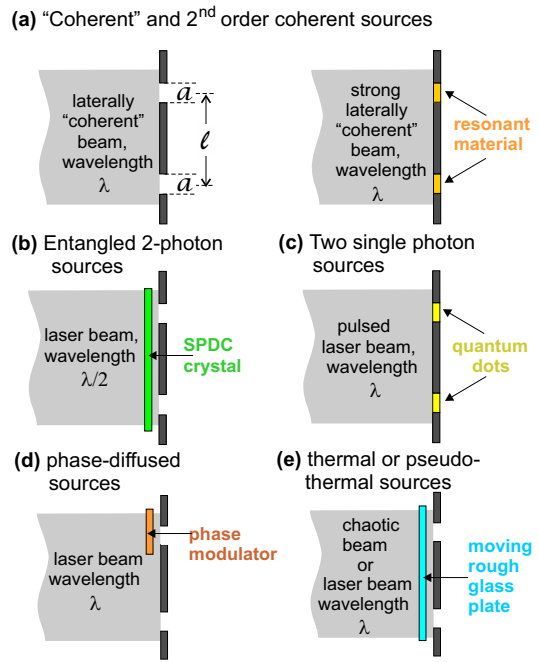}
\caption[]{Experimental schemes of preparing different double-slit-like sources representing different quantum states, as discussed in the text.
}
\label{Fig:2-photon-sources}
\end{figure}

The advent of the laser has allowed the creation of different quantum states of light which are described by quantum optics \cite{loudon,scully-zubairy,Ou:2007,Shih:2011,Ou:2017}. In Fig.\,\ref{Fig:2-photon-sources} we present different schemes that have been used to prepare double-slit-like sources. The reason for the shown order will become clear later when the cases will be linked to the evolution of their respective diffraction patterns from first to second order.

Fig.\,\ref{Fig:2-photon-sources}\,(a) shows the cases of ``coherent'' illumination of the double slits by a conventional source that has been made (first order) ``coherent'' by use of a monochromator and pinhole or by a laser which is higher order coherent. The first and second order diffraction patterns have been studied by  Shimizu \emph{et al.}  \cite{shimizu:2006} using either a monochromatized halogen lamp  or a Ti:sapphire laser to illuminate the slits.

On the right of the same figure we show the particular case where a strong incident near-transform-limited pulse, whose width is approximately equal to the photon coherence time,  is tuned to a well defined atomic resonance in a thin film \cite{wu:2016}. In the so-prepared source, absorption is compensated by stimulated emission \cite{stohr-scherz:15}, producing a second order coherent source \cite{stohr-AOP}.

In Fig.\,\ref{Fig:2-photon-sources}\,(b), a laser is used to illuminate a suitable thin crystal that through spontaneous parametric down-conversion produces two spatially entangled photons, each with half the incident photon energy \cite{Klyshko:88}. The ``entangled biphoton'' diffraction case has been extensively studied in the literature \cite{Shih:2011,liu:2010,stohr-AOP}.

In Fig.\,\ref{Fig:2-photon-sources}\,(c) two \emph{single photons} are simultaneously emitted from two quantum sources. This case can be implemented by use of a  laser pulse that triggers two quantum dots or two trapped atoms or ions to simultaneously emit single photons \cite{santori:2002,patel:2010,flagg:2010,neuzner:2016}.

In Fig.\,\ref{Fig:2-photon-sources}\,(d) the slits are illuminated by \emph{phase-diffused light}, implemented by Liu \emph{et al.} \,\cite{liu:2010} by splitting a coherent laser beam and modulating the phase of one of the beams with a vibrating mirror. The intensity falling onto the two slits is kept \emph{constant} so that the photons with random phases still obey \emph{Poisson} counting statistics.

Finally, as shown in Fig.\,\ref{Fig:2-photon-sources}\,(e), the slits may be illuminated by \emph{chaotic} light  produced by a thermal source \cite{zhai:2005}.  In practice, it is convenient to use higher intensity ``pseudo-thermal'' light generated by shining a laser on a rotating ground glass plate \cite{martienssen:64}. Such light exhibits both phase and intensity fluctuations with \emph{Bose-Einstein} counting statistics \cite{arecchi:65}. It has been employed for double slit diffraction by several groups \cite{scarcelli:2004, xiong:2005, Zhou:2017}.

The cases shown in Fig.\,\ref{Fig:2-photon-sources} represent experimental schemes that generate specific quantum states of light whose  characteristic first and second order diffraction patterns have been reported in the cited literature. In the following we shall illustrate how the observed diffraction patterns follow from the photon-based formulation of diffraction within QED, which directly links quantum states with their characteristic encoded diffraction signatures.

The presented formulation of quantum diffraction is general up to second order in QED and covers the infinite number of quantum states of light within QED. These quantum states give rise to the complex light behavior typically described by the broad and difficult concept of ``partial coherence'' in statistical optics \cite{Born-Wolf,Goodman-SO}. The quantum formulation allows the specification of this concept by directly defining coherence and diffraction through that of individual quantum states. These states span the entire range between the limiting cases of \emph{coherent states} represented by a Poisson distribution and \emph{chaotic states} associated with a Bose-Einstein distribution.

\section{The Formulation of Quantum Diffraction}
\label{SS:quantum-diff}

\begin{figure}[!h]
\centering
\includegraphics[width=0.75\columnwidth]{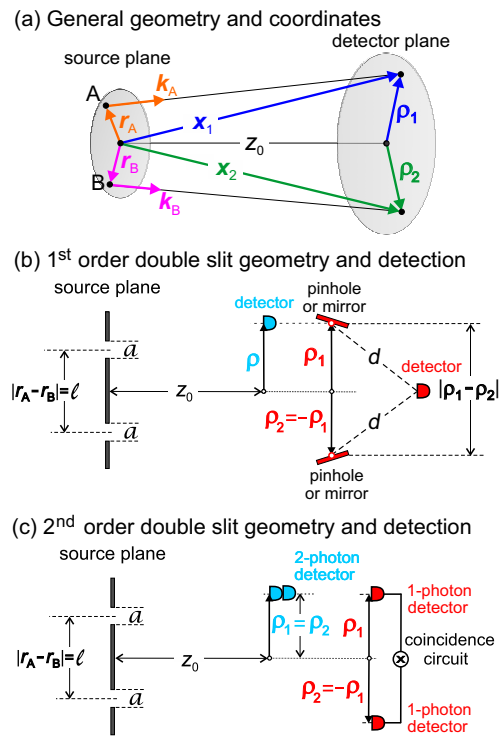}
\caption[]{(a) Assumed geometry and coordinates of photon propagation from points $\vec r_A$ and $\vec r_B$ in the source plane to   distant  points with  coordinates $\vec \rho_i$ or $\vec x_i$ ($i=1,2$) in the detector plane.
(b) In first order, photons emitted from either slit are detected at a \emph{single} point using the scenarios shown in blue and red. A photon detector is either scanned as a function of its separation $\rho $ from the optical axis or a detector on the optical axis detects the probability that photons have taken paths through points $\vec \rho_2=-\vec \rho_1$, defined by scannable mirrors or pinholes. The detection probability is given by (\ref{Eq:Det-prob-O}) with $O=1$. (c) In second order, one measures the coincident arrival of two photons at a single or two points by use of the blue and red detection schemes. The detection probability is given by (\ref{Eq:Det-prob-O}) with $O=2$.  }
\label{Fig:geometry}
\end{figure}

The experimental geometry for Young's double slit experiment is illustrated in Fig.\,\ref{Fig:geometry}\,(a) with identification of the relevant coordinates.  The photons emerging from the two source points A and B, which may be expanded into slits, are detected at a single point or two points in a distant detector plane.

The diffraction pattern is determined by  the \emph{spatial} variations in the number of detected quasi-monochromatic photons. The assumption of monochromaticity allows the reduction of the general space-time description of photons in QED to the spatial domain only, since the width of the photon energy distribution defines a time interval of photon arrival, given by the \emph{coherence time} $\tau$ of the photons. This is taken into account in the design of diffraction experiments, as illustrated in Fig.\,\ref{Fig:geometry}\,(b) and (c). The first order diffraction pattern is determined by the arrival probability of photons within the time increment $\tau$ at a given detection point, while the second order pattern is given by the coincident arrival probability of two photons within $\tau$ at a single or two detection points.

The first and second order diffraction patterns correspond to the lowest orders, $O=1,2$, of perturbation in QED. As shown below, they may be calculated by use of Feynman's probability amplitude formulation \cite{liu:2010,stohr-AOP} or equivalently by means of Glauber's order-dependent correlation functions \cite{glauber:65,scully-zubairy}. In the latter formulation, which underlies modern quantum optics, the patterns are determined by photon creation at source points, photon propagation via straight paths to detection points, where the photons are destroyed.  Photon birth and destruction are treated on equal footing through quantum mechanical operators. The patterns may be written in the simple form,
\begin{eqnarray}
\left \langle \Phi_\mathrm{s} |\mathbf  P^{(O)}(\vec \rho_1, \vec \rho_2  )|\Phi_\mathrm{s} \right \rangle = P_O \, \boldsymbol {{\cal G}}^{(O)}  (\vec \rho_1, \vec \rho_2  )
\label{Eq:Det-prob-O}
\end{eqnarray}
Here $\mathbf  P^{(O)}(\vec \rho_1, \vec \rho_2  )$ is the detection probability operator, given by the conventional quantum optical correlation function of order $O$ written in operator form \cite{glauber:65,scully-zubairy}. The triangular brackets denote the quantum mechanical expectation value, evaluated for a given quantum state $|\Phi_\mathrm{s}\rangle$ created at the source. This formulation establishes the direct correspondence between a created quantum state and its first and second order diffraction patterns.

As indicated on the right of (\ref{Eq:Det-prob-O}), the diffraction pattern can be conveniently written in terms of a position-dependent shape function $\boldsymbol {{\cal G}}^{(O)}  (\vec \rho_1, \vec \rho_2  )$ and an overall scaling factor $P_O$ that is determined by conservation of emitted and detected photons.

\subsection{First Order Diffraction}

The first order probability operator $ \mathbf  P^{(1)}(\vec \rho_1, \vec \rho_2  )$ in (\ref{Eq:Det-prob-O}) may be expressed either in the coordinates of the detection points $\vec \rho_i$ or $\vec x_i$ defined in Fig.\,\ref{Fig:geometry}\,(a). For large source-detector separation $z_0$ and wavevectors of equal magnitude  $k=|\vec k_\mathrm{A}|=|\vec k_\mathrm{B}|=2\pi/\lambda$, the conversion of the   coordinates is given by the simple relation $\vec k_\mathrm{I} \cdot \vec x_i= k (z_0- \vec r_\mathrm{I} \cdot \vec \rho_i/z_0)$, where  I=A,B  and $i\!=\!1,2$ \cite{stohr-AOP}.

In the coordinates $(\vec x_i,\vec k_\mathrm{I})$ of Fig.\,\ref{Fig:geometry}\,(a), the expectation value of the first order probability operator  $ \mathbf  P^{(1)}(\vec x_1, \vec x_2  )$ is given by the absolute value squared of the \emph{total photon probability amplitude} at position $\vec x_i$, denoted $\Phi(\vec{x}_i )$, according to $\langle  \mathbf  P^{(1)}(\vec x_1, \vec x_2  ) \rangle = \langle \Phi^*(\vec{x}_1)\Phi(\vec{x}_2 )  \rangle$, where \cite{scully-zubairy}
\begin{eqnarray}
\Phi^*(\vec{x}_1 ) \! &=& \!\!  \frac{1}{\sqrt{2}}    \left( \mathbf  a_{\vec{k}_\mathrm{A}}^{\dagger}     \mathrm{e}^{ - \mathrm{i} \vec{k}_\mathrm{A} \cdot \vec{x}_1 } + \mathbf  a_{\vec{k}_\mathrm{B}}^{\dagger}     \mathrm{e}^{- \mathrm{i} \vec{k}_\mathrm{B} \cdot \vec{x}_1 } \right)
\nonumber \\
\Phi(\vec{x}_2 ) \! &=& \!\!    \frac{1}{\sqrt{2}}    \left( \mathbf  a_{\vec{k}_\mathrm{A}}   \mathrm{e}^{ \mathrm{i} \vec{k}_\mathrm{A} \cdot \vec{x}_2} + \mathbf  a_{ \vec{k}_\mathrm{B}}    \mathrm{e}^{ \mathrm{i} \vec{k}_\mathrm{B}  \cdot \vec{x}_2} \right)
\label{Eq:Glauber-1-P-amplitudes}
\end{eqnarray}
These probability amplitudes directly correspond to the \emph{single photon} probability amplitudes in Feynman's formulation \cite{stohr-AOP},
\begin{eqnarray}
\Phi^*(\vec{x}_1 ) \! &=& \!\!  \frac{1}{\sqrt{2}}   \left( \mathrm{e}^{ -\mathrm {i} \alpha  }    \mathrm{e}^{ - \mathrm{i} \vec{k}_\mathrm{A} \cdot \vec{x}_1 } + \mathrm{e}^{- \mathrm {i}   \beta}    \mathrm{e}^{- \mathrm{i} \vec{k}_\mathrm{B} \cdot \vec{x}_1 } \right)
\nonumber \\
\Phi(\vec{x}_2 ) \! &=& \!\!    \frac{1}{\sqrt{2}}   \left(  \mathrm{e}^{  \mathrm {i} \alpha  }   \mathrm{e}^{ \mathrm{i} \vec{k}_\mathrm{A} \cdot \vec{x}_2} +  \mathrm{e}^{  \mathrm {i} \beta  }  \mathrm{e}^{ \mathrm{i} \vec{k}_\mathrm{B}  \cdot \vec{x}_2} \right)
\label{Eq:Feynman-1-P-amplitudes}
\end{eqnarray}
In first order QED, the abstract creation and annihilation operators are simply replaced by phase factors, containing birth phases $\alpha$ and $\beta$.

The \emph{first order} diffraction pattern may then be written as the dimensionless detection probability,
\begin{eqnarray}
\left \langle \Phi_\mathrm{s} |\mathbf  P^{(1)}(\vec x_1, \vec x_2  ) |\Phi_\mathrm{s} \right\rangle\!=\!\frac{1}{2}  \bigg \langle  \Phi_\mathrm{s} |\mathbf  X    +   \mathbf  Y   |\Phi_\mathrm{s}  \bigg \rangle
\label{Eq:SM-1-photon-probability}
\end{eqnarray}
where the expectation value is evaluated for a given quantum state $|\Phi_\mathrm{s}\rangle$ created in the source.
The operators $\mathbf  X$ and $\mathbf  Y $ are products of the amplitudes (\ref{Eq:Glauber-1-P-amplitudes}) given by,
 \begin{eqnarray}
\mathbf  X  = \mathbf  a_{\vec{k}_\mathrm{A}}^{\dagger} \mathbf  a_{\vec{k}_\mathrm{A}}   \mathrm{e}^{  \mathrm {i} \vec{k}_\mathrm{A} \cdot (\vec{x}_2- \vec{x}_1) }+ \mathbf  a_{\vec{k}_\mathrm{B}}^{\dagger}  \mathbf  a_{\vec{k}_\mathrm{B}}\mathrm{e}^{ \mathrm {i} \vec{k}_\mathrm{B} \cdot (\vec{x}_2 -   \vec{x}_1)}
\label{Eq:SM-term-X}
\end{eqnarray}
\begin{eqnarray}
\mathbf  Y = \mathbf  a_{\vec{k}_\mathrm{A}}^{\dagger} \mathbf  a_{\vec{k}_\mathrm{B}}   \mathrm{e}^{   \mathrm {i} (\vec{k}_\mathrm{B} \cdot \vec{x}_2 -\vec{k}_\mathrm{A} \cdot \vec{x}_1) } \!+ \!\mathbf  a_{\vec{k}_\mathrm{B}}^{\dagger} \mathbf  a_{\vec{k}_\mathrm{A}}    \mathrm{e}^{  \mathrm {i}( \vec{k}_\mathrm{A} \cdot \vec{x}_2 - \vec{k}_\mathrm{B} \cdot \vec{x}_1) }
\label{Eq:SM-term-Y}
\end{eqnarray}
where  $\mathbf  X$   contains only pairs of creation and destruction operators with the same wavevector mode, while $\mathbf  Y$ contains pairs of creation and destruction operators in different modes.

\subsection{Second Order Diffraction}

The \emph{second order} detection \emph{probability} is given by $\langle  \mathbf  P^{(2)}(\vec x_1, \vec x_2  ) \rangle = \langle |\Psi (\vec x_1,\vec x_2|^2 \rangle$ where  the second order \emph{photon probability amplitude} is given by \cite{scully-zubairy},
\begin{eqnarray}
&\,& \hspace*{-20pt} \Psi (\vec x_1,\vec x_2)
\nonumber \\
&=&\! \frac{1}{2}\bigg\{
 \mathbf a_{\vec{k}_\mathrm{A}}    \mathbf a_{\vec{k}_\mathrm{A}}   \mathrm{e}^{   \mathrm{i} \vec{k}_\mathrm{A} \cdot \vec{x}_1 }  \mathrm{e}^{  \mathrm{i} \vec{k}_\mathrm{A}\cdot \vec{x}_2 }
\!+\!
\mathbf a_{\vec{k}_\mathrm{A}}   \mathbf  a_{ \vec{k}_\mathrm{B}}    \mathrm{e}^{  \mathrm{i} \vec{k}_\mathrm{A}\cdot \vec{x}_2 } \mathrm{e}^{ \mathrm{i} \vec{k}_\mathrm{B} \cdot \vec{x}_1 }
\nonumber \\
  &\,&   +
\mathbf  a_{ \vec{k}_\mathrm{B}}  \mathbf a_{\vec{k}_\mathrm{A}}   \mathrm{e}^{  \mathrm{i} \vec{k}_\mathrm{A}\cdot \vec{x}_1 }  \mathrm{e}^{  \mathrm{i} \vec{k}_\mathrm{B} \cdot \vec{x}_2 }
\!+\!
\mathbf  a_{ \vec{k}_\mathrm{B}}  \mathbf  a_{ \vec{k}_\mathrm{B}}   \mathrm{e}^{  \mathrm{i} \vec{k}_\mathrm{B} \cdot \vec{x}_1 } \mathrm{e}^{  \mathrm{i} \vec{k}_\mathrm{B} \cdot \vec{x}_2 } \bigg\}~
\label{Eq:G2-Glauber-wavefunction}
\end{eqnarray}
Here we have used the commutation relation  $\mathbf  a_{ \vec{k}_\mathrm{A} }  \mathbf  a_{ \vec{k}_\mathrm{B}}\!= \!\mathbf  a_{ \vec{k}_\mathrm{B}}   \mathbf  a_{ \vec{k}_\mathrm{A}}$, which similarly holds for the creation operators \cite{glauber:65}.
As for the first order case, Glauber's probability amplitude expression (\ref{Eq:G2-Glauber-wavefunction}) is just the operator form of Feynman's formulation, where the operators are replaced by phase factors containing the birth phases  of photons at the two source points according to \cite{liu:2010,stohr-AOP},
\begin{eqnarray}
&\,&\hspace*{-28pt} \Psi  (\vec x_1,\vec x_2)
\nonumber \\
 &=& \hspace*{-5pt} \frac{1}{2}\bigg\{
\mathrm{e}^{ \mathrm {i} \left[\alpha_1 +\alpha'_1 \right]} \,  \mathrm{e}^{    \mathrm{i} \vec{k}_\mathrm{A} \cdot \vec{x}_1 }  \mathrm{e}^{  \mathrm{i} \vec{k}_\mathrm{A}\cdot \vec{x}_2 }
\!+\!
\mathrm{e}^{ \mathrm {i} \left[\alpha'_0  + \beta_0 \right]} \,  \mathrm{e}^{  \mathrm{i} \vec{k}_\mathrm{A}\cdot \vec{x}_2 } \mathrm{e}^{ \mathrm{i} \vec{k}_\mathrm{B} \cdot \vec{x}_1 }
\nonumber \\
  &\,& \hspace*{-7pt}  +
\mathrm{e}^{ \mathrm {i}  \left[\alpha_0  +\beta'_0  \right]} \,\mathrm{e}^{  \mathrm{i} \vec{k}_\mathrm{A}\cdot \vec{x}_1 }  \mathrm{e}^{  \mathrm{i} \vec{k}_\mathrm{B} \cdot \vec{x}_2 }
\!+\!
\mathrm{e}^{ \mathrm {i} \left[\beta_1  +\beta'_1 \right]} \, \mathrm{e}^{   \mathrm{i} \vec{k}_\mathrm{B} \cdot \vec{x}_1 } \mathrm{e}^{  \mathrm{i} \vec{k}_\mathrm{B} \cdot \vec{x}_2 } \!\bigg\}~~
\label{Eq:G2-Feynman-wavefunction}
\end{eqnarray}
The two formulations are seen to be formally equivalent.

Feynman's formulation is based on specific rules \cite{feynman-QED} how, for a given case, multiple-photon probability amplitudes are constructed from 1-photon amplitudes, either by addition (``alternative'' photon paths) or multiplication (``concomitant'' photon paths) \cite{stohr-AOP}. This procedure becomes increasingly complicated beyond the 2-photon case.

Glauber's abstract formulation can be more easily extended  to many photons \cite{glauber:65,scully-zubairy}, and for this reason underlies modern quantum optics. Most importantly, one does not have to worry about birth phases since they are effectively determined by the matrix elements of the operator $\mathbf  P^{(2)}(\vec x_1, \vec x_2)$, i.e. the expectation value calculated with different quantum states of light created in the source. This leads to the link of quantum states of light and their diffraction patterns. This link demonstrated in the present paper up to second order in QED, may also be extended to orders $O>2$, describing  higher-order multi-photon interference, which can be measured today with multi-element single-photon detectors \cite{liu:2009,stevens:2010}.

The \emph{second order} detection \emph{probability} may be written as,
\begin{eqnarray}
\left \langle   \Phi_\mathrm{s} | \mathbf  P^{(2)}(\vec x_1, \vec x_2  ) | \Phi_\mathrm{s}  \right\rangle = \frac{1}{4}  \bigg \langle  \Phi_\mathrm{s}  | \mathbf  A + \mathbf  B + \mathbf  C + \mathbf D  |\Phi_\mathrm{s}  \bigg  \rangle
\label{Eq:SM-2-photon-probability}
\end{eqnarray}
where the quantum state $|\Phi_\mathrm{s}\rangle$ is created in the source. The four operators $\mathbf A ,\mathbf  B,\mathbf C ,\mathbf  D  $  are explicitly given by,
\begin{eqnarray}
\hspace*{-10pt}\mathbf A \! &=&\!\!
\mathbf a_{\vec{k}_\mathrm{A}}^{\dagger}\! \mathbf  a_{ \vec{k}_\mathrm{B}}^{\dagger}  \!
\mathbf a_{\vec{k}_\mathrm{A}} \!   \mathbf  a_{ \vec{k}_\mathrm{B}}
 +
\mathbf a_{\vec{k}_\mathrm{A}}^{\dagger}  \!  \mathbf  a_{ \vec{k}_\mathrm{B}}^{\dagger}
\mathbf a_{\vec{k}_\mathrm{A}} \!  \mathbf  a_{ \vec{k}_\mathrm{B}}
\nonumber \\
&\,& \! +
\mathbf a_{\vec{k}_\mathrm{A}}^{\dagger}  \!  \mathbf  a_{ \vec{k}_\mathrm{B}}^{\dagger}\!
\mathbf a_{\vec{k}_\mathrm{A}}   \mathbf  a_{ \vec{k}_\mathrm{B}}
\mathrm{e}^{ - \mathrm{i} (\vec{k}_\mathrm{A}-\vec{k}_\mathrm{B})\cdot(\vec{x}_1 - \vec{x}_2) }
\nonumber \\
&\,& \! +
 \mathbf a_{\vec{k}_\mathrm{A}}^{\dagger}\! \mathbf  a_{ \vec{k}_\mathrm{B}}^{\dagger} \!  \mathbf a_{\vec{k}_\mathrm{A}} \!
\mathbf  a_{ \vec{k}_\mathrm{B}}
\mathrm{e}^{  \mathrm{i} (\vec{k}_\mathrm{A} -\vec{k}_\mathrm{B}) \cdot (\vec{x}_1 - \vec{x}_2)}
\nonumber \\
 &=&\!\!
4\, \mathbf a_{\vec{k}_\mathrm{A}}^{\dagger}\! \mathbf  a_{ \vec{k}_\mathrm{B}}^{\dagger}  \!
\mathbf a_{\vec{k}_\mathrm{A}} \!   \mathbf  a_{ \vec{k}_\mathrm{B}}
 \cos^2\!\left[\frac{1}{2} (\vec{k}_\mathrm{A}-\vec{k}_\mathrm{B})\cdot(\vec{x}_1 - \vec{x}_2)\right]
\label{Eq:SM-term-A}
\end{eqnarray}
\begin{eqnarray}
\mathbf B\! &=&\!\! \mathbf a_{\vec{k}_\mathrm{A}}^{\dagger} \!    \mathbf a_{\vec{k}_\mathrm{A}}^{\dagger} \!
\mathbf a_{\vec{k}_\mathrm{A}} \! \mathbf a_{\vec{k}_\mathrm{A}}
 +
\mathbf  a_{ \vec{k}_\mathrm{B}}^{\dagger} \!  \mathbf  a_{ \vec{k}_\mathrm{B}}^{\dagger} \!
\mathbf  a_{ \vec{k}_\mathrm{B}} \!  \mathbf  a_{ \vec{k}_\mathrm{B}}
\nonumber \\
&\,& +
\mathbf a_{\vec{k}_\mathrm{A}}^{\dagger} \!    \mathbf a_{\vec{k}_\mathrm{A}}^{\dagger} \!
\mathbf  a_{ \vec{k}_\mathrm{B}} \! \mathbf  a_{ \vec{k}_\mathrm{B}}
\mathrm{e}^{ - \mathrm{i} (\vec{k}_\mathrm{A}-\vec{k}_\mathrm{B})\cdot (\vec{x}_1 +\vec{x}_2) }
\nonumber \\
&\,& \!+
\mathbf  a_{ \vec{k}_\mathrm{B}}^{\dagger} \!  \mathbf  a_{ \vec{k}_\mathrm{B}}^{\dagger}\!
\mathbf a_{\vec{k}_\mathrm{A}}  \!   \mathbf a_{\vec{k}_\mathrm{A}}
\mathrm{e}^{  \mathrm{i}(\vec{k}_\mathrm{A}-\vec{k}_\mathrm{B}) \cdot (\vec{x}_1 + \vec{x}_2) }
\label{Eq:SM-term-B}
\end{eqnarray}
\begin{eqnarray}
\hspace*{-20pt}\mathbf  C\! &=&\!\!  \left[\mathbf a_{\vec{k}_\mathrm{A}}^{\dagger} \!    \mathbf a_{\vec{k}_\mathrm{A}}^{\dagger}\!
\mathbf a_{\vec{k}_\mathrm{A}} \! \mathbf  a_{ \vec{k}_\mathrm{B}}
\!+\!
\mathbf a_{\vec{k}_\mathrm{A}}^{\dagger}  \!  \mathbf  a_{ \vec{k}_\mathrm{B}}^{\dagger}\!
\mathbf  a_{ \vec{k}_\mathrm{B}} \!  \mathbf  a_{ \vec{k}_\mathrm{B}}\right]\!
\mathrm{e}^{ - \mathrm{i} (\vec{k}_\mathrm{A}-\vec{k}_\mathrm{B})\cdot \vec{x}_1 }
\nonumber \\
&\,& \! +
\left[\mathbf  a_{ \vec{k}_\mathrm{B}}^{\dagger}  \!  \mathbf a_{\vec{k}_\mathrm{A}}^{\dagger}\!
\mathbf a_{\vec{k}_\mathrm{A}}  \!   \mathbf a_{\vec{k}_\mathrm{A}}
\!+\!
\mathbf  a_{ \vec{k}_\mathrm{B}}^{\dagger} \!  \mathbf  a_{ \vec{k}_\mathrm{B}}^{\dagger}\!
\mathbf  a_{ \vec{k}_\mathrm{B}} \!  \mathbf a_{\vec{k}_\mathrm{A}}\right] \!
\mathrm{e}^{\mathrm{i} (\vec{k}_\mathrm{A}-\vec{k}_\mathrm{B}) \cdot \vec{x}_1 }~
\label{Eq:SM-term-C}
\end{eqnarray}
\begin{eqnarray}
\hspace*{-20pt}\mathbf  D\! &=&\!\!  \left[\mathbf a_{\vec{k}_\mathrm{A}}^{\dagger} \!    \mathbf a_{\vec{k}_\mathrm{A}}^{\dagger}\! \mathbf a_{\vec{k}_\mathrm{A}} \!
\mathbf  a_{ \vec{k}_\mathrm{B}}
\!+\!
a_{\vec{k}_\mathrm{A}}^{\dagger} \! \mathbf  a_{ \vec{k}_\mathrm{B}}^{\dagger}  \!  \mathbf \mathbf a_{ \vec{k}_\mathrm{B}} \!  \mathbf  a_{ \vec{k}_\mathrm{B}}\right] \!
\mathrm{e}^{ - \mathrm{i} (\vec{k}_\mathrm{A}-\vec{k}_\mathrm{B})\cdot \vec{x}_2 }
\nonumber \\
&\,& \! +
\left[\mathbf  a_{ \vec{k}_\mathrm{B}}^{\dagger} \! \mathbf a_{\vec{k}_\mathrm{A}}^{\dagger}  \!
\mathbf a_{\vec{k}_\mathrm{A}}  \!   \mathbf a_{\vec{k}_\mathrm{A}}
\!+\!
\mathbf  a_{ \vec{k}_\mathrm{B}}^{\dagger} \!  \mathbf  a_{ \vec{k}_\mathrm{B}}^{\dagger}\!
 \mathbf  a_{ \vec{k}_\mathrm{B}} \! \mathbf a_{\vec{k}_\mathrm{A}}\right]\!
\mathrm{e}^{   \mathrm{i}(\vec{k}_\mathrm{A}- \vec{k}_\mathrm{B}) \cdot \vec{x}_2 }~
\label{Eq:SM-term-D}
\end{eqnarray}
These expressions consist of normally ordered products of two creation and two destruction operators which obey the quantum mechanical commutation relations \cite{Dirac-book}. The order of two adjacent creation or destruction operators may be switched, but not the order in a  pair formed by single creation and destruction operators \cite{glauber:65}.

\subsection{Order-Dependent Degree of Coherence}

In quantum optics, coherence is characterized by a  \emph{degree of spatial coherence}  $g^{(O)}(\vec x_1,\vec x_2)$ which depends on the order $O$ of perturbation in QED. The degrees of first ($O\!=\!1$) and second ($O\!=\!2$) order spatial coherence are defined as \cite{glauber:65,loudon},
\begin{eqnarray}
g^{(O)}(\vec x_1,\vec x_2) =\frac{\left \langle \mathbf  P^{(O)}(\vec x_1, \vec x_2  )  \right\rangle}
{\left[\left \langle \mathbf  P^{(1)}(\vec x_1, \vec x_1  )  \right\rangle \left \langle \mathbf  P^{(1)}(\vec x_2, \vec x_2  )  \right\rangle\right]^{O/2}}
\label{Eq:SM-Degree-of-coherence}
\end{eqnarray}

For the \emph{coherent cases}, the numerators in $ g^{(O)}(\vec x_1,\vec x_2)$ factor into the denominators, so that the diffraction fine structure contained in both the numerators and denominators is normalized out, yielding a constant. It is then convenient to plot the normalized diffraction pattern  $\boldsymbol {{\cal G}}^{(O)}  (\vec x_1, \vec x_2  )$ defined through (\ref{Eq:Det-prob-O}) which preserves the characteristic diffraction structure. We shall utilize both complementary formulations in the present paper.

\section{The Quantum States of Light}

The first and second order diffraction patterns, defined by (\ref{Eq:Det-prob-O}) with $O=1,2$, are determined by quantum states involving two wavevector modes  $\vec k_\mathrm{A}$ and $\vec k_\mathrm{B}$ as defined in  Fig.\,\ref{Fig:geometry}\,(a).  In the following we will switch to the shorter and more convenient notation $\vec k=\vec k_\mathrm{A}$ and $\vec k'=\vec k_\mathrm{B}$.

The 2-mode quantum states produced in the cases shown in Figs.\,\ref{Fig:2-photon-sources} involve different numbers of photons. In general, we distinguish  \emph{collective states} which contain an \emph{average} number of photons per mode from states that contain a \emph{specific} number of photons per mode. We first discuss the 2-mode multi-photon collective states associated with Figs.\,\ref{Fig:2-photon-sources}\,(a), (d) and (e) and their decomposition into probability distributions of substates. Their specific 2-photon substates are then linked to the central two cases in Figs.\,\ref{Fig:2-photon-sources}\,(b) and (c), which involve only two photons.

\subsection{2-Mode Collective Quantum States}

The 2-mode collective \emph{coherent state} produced in Fig.\,\ref{Fig:2-photon-sources}\,(a) and the \emph{ phase-diffused coherent state} in Fig.\,\ref{Fig:2-photon-sources}\,(d) are constructed from two single mode coherent states of the form \cite{loudon},
\begin{equation}
|\alpha \rangle_k =
\sum_{m=0}^\infty \,  \frac{ \alpha_k^m}{\mathrm{e}^{|\alpha_k|^2/2}\,\sqrt {m! }} \, |m \rangle_k
 \label{Eq:SM-coh-1-mode-state}
\end{equation}
Here $\alpha_k$ is a complex number and the coherent state contains an \emph{average number of photons}  $\langle n \rangle_k=|\alpha_k|^2$ in  the mode $k$, distributed in a Poisson distribution around the average value $|\alpha_k|^2$.  In general, the two modes may contain different numbers of photons and have different phases. In the following we shall assume that both modes contain the same \emph{average} number of photons per mode, i.e. $|\alpha|^2=\langle n\rangle =\langle n\rangle_{k}=\langle n\rangle_{k'}$. We then obtain with $\alpha_{k} \!=\!|\alpha | \mathrm{e}^{\mathrm{i} \phi_k }$  and $\alpha_{k'} \!=\!|\alpha | \mathrm{e}^{\mathrm{i} \phi_{k'} }$ the following general expression for a 2-mode collective ``coherent'' state,
\begin{eqnarray}
|\alpha\rangle_k  |\alpha \rangle_{k'} = \frac{1}{\mathrm{e}^{|\alpha|^2 }}
\sum_{n=0}^\infty \sum_{m=0}^\infty  \frac{|\alpha|^{n+m}\,\mathrm{e}^{\mathrm{i}(n\phi_{k}+m\phi_{k'}) } }{\sqrt{n!\, m!}}   \, |n \rangle_k  \, |m \rangle_{k'}
\hspace*{-15pt}
\nonumber \\
\label{Eq:SM-2-mode-coh-state-general}
\end{eqnarray}

The \emph{ 2-mode chaotic state} associated with Fig.\,\ref{Fig:2-photon-sources}\,(e) is constructed from two single mode chaotic states $|\beta\rangle_k$ given by \cite{loudon}
\begin{equation}
 |\beta \rangle_k = \sum_{m=0}^\infty \sqrt {\frac{\langle n \rangle^{m}}{\left( 1 + \langle n
\rangle \right)^{1+m}}} \, |m \rangle_k
 \label{Eq:SM-cha-1-mode-state}
\end{equation}
It contains an average number of $\langle n \rangle$ photons per mode in the form of a Bose-Einstein distribution. The \emph{2-mode chaotic state} containing an \emph{average number} $\langle n \rangle$ of photons in each mode is given by
\begin{eqnarray}
 |\beta \rangle_k |\beta \rangle_{k'}
= \sum_{\ell=0}^\infty \sum_{m=0}^\infty \! \mathrm{e}^{\mathrm{i}( \phi_\ell +\phi_m)} \sqrt{ \frac{\langle n  \rangle^{\ell} \langle n  \rangle^m}{ ( 1\! + \!\langle n
\rangle  )^{\ell+m + 2} }} \, |\ell\rangle_{k}|m \rangle_{k'}
\hspace*{-15pt}
\nonumber \\
\label{Eq:SM-cha-2-mode-state-0}
\end{eqnarray}

The states (\ref{Eq:SM-2-mode-coh-state-general}) and (\ref{Eq:SM-cha-2-mode-state-0}) are \emph{collective} 2-mode quantum states which contain the same \emph{average} number of photons per mode, $\langle n\rangle=|\alpha|^2$. We now show that they may be written as a linear combination of 2-mode  \emph{substates} that contain \emph{specific numbers} of photons $N=0,1,2,3...\infty$ with probability distributions around the mean value $2\langle n \rangle=N$.

\subsection{The Collective Coherent State and its Substates}

The \emph{2-mode coherent state} describes the case shown in Fig.\,\ref{Fig:2-photon-sources}\,(a) where the two slits are illuminated by the same average number of photons in both modes and the two modes have the same phases. We then  have $\alpha_{k}\!=\!\alpha_{k'} \!=\!|\alpha | \mathrm{e}^{\mathrm{i} \phi }$ and the general expression (\ref{Eq:SM-2-mode-coh-state-general}) can be written in the form,
\begin{eqnarray}
&\,& \hspace*{-15pt} |\Phi_\mathrm{coh} \rangle_{k,k'}
\nonumber \\
&=&\! \!\frac{1}{\mathrm{e}^{|\alpha|^2 }} \! \sum_{N=0}^\infty |\alpha|^{N}\,  \mathrm{e}^{\mathrm{i} N \phi}  \!
\sum_{m =0}^N \! \frac{1}{\sqrt{m! \,(N\!-\!m)!}}  \, |m\rangle_k  \, | N\!-\!m  \rangle_{k'}~~~
\hspace*{-15pt}
\nonumber \\
\label{Eq:SM-coh-2-mode-state}
\end{eqnarray}
The state is composed of \emph{binomial substates} which contain \emph{specific numbers} $N$ of photons that are distributed according to a \emph{Poisson} probability distribution around the average value $2\langle n \rangle$. We can write,
\begin{equation}
|\Phi_\mathrm{coh} \rangle_{k,k'}=    \sum_{N=0}^\infty   \underbrace{\frac{2^{N/2}\,\alpha^N } {\sqrt {N!}\, \mathrm{e}^{ |\alpha|^2}} }_{\mbox{$ c^\alpha_N$}}\,  |\phi_\mathrm{cohN}\rangle_{k,k'}
\label{Eq:coh-state-cN}
\end{equation}
The complex coefficients  $ c^\alpha_N$  fulfill the normalization $\sum_{N=0}^\infty   |c^\alpha_N|^2 =1$ and weigh the contributions of the  \emph{ binomial substates}  $|\phi_\mathrm{cohN}\rangle_{k,k'}$ which contain $N$ photons and are given by
\begin{eqnarray}
\hspace*{-10pt}|\phi_\mathrm{cohN}\rangle_{k,k'}\!  & = & \!  \frac{1}{2^{N/2}} \sum_{m =0}^N \sqrt{\frac{N !}{m! \,(N\!-\!m)!}}  ~ |m\rangle_k
 \, | N\!-\!m  \rangle_{k'} \nonumber \\
\label{Eq:SM-coh-N-bin-substates}
\end{eqnarray}
The states  fulfill the normalization $\langle \Phi_\mathrm{coh} |\Phi_\mathrm{coh} \rangle=\langle \phi_\mathrm{cohN} |\phi_\mathrm{cohN} \rangle=1$.

\subsection{The Collective Phase-Diffused Coherent State and its Substates}

The \emph{phase-diffused laser light} encountered for the case in   Fig.\,\ref{Fig:2-photon-sources}\,(d) corresponds to random phases between the two modes in (\ref{Eq:SM-2-mode-coh-state-general}). Since only the relative phase between the two modes is important, we may set $\phi_{k}=0$, and denote the relative phaseshift as $\varphi =\phi_{k'} $ to obtain,
\begin{eqnarray}
&\,&\hspace*{-20pt} |\Phi_\mathrm{dif} \rangle_{k,k'}
\nonumber \\
&=&\!\! \frac{1}{\mathrm{e}^{|\alpha|^2 }}
\sum_{N=0}^\infty |\alpha|^{N} \sum_{m=0}^N \frac{\mathrm{e}^{\mathrm{i} (N-m) \varphi}}{\sqrt{m!\, (N-m)!}}   \, |m \rangle_k  \, |N\!-\!m \rangle_{k'} ~~~
\hspace*{-15pt}
\nonumber \\
\label{Eq:SM-dif-2-mode-coh-state}
\end{eqnarray}
This can be written in terms of substates containing a specific number of $N$ photons with probabilities in form of a \emph{Poisson} distribution around the average value $2\langle n \rangle$ according to,
\begin{eqnarray}
|\Phi_\mathrm{dif} \rangle_{k,k'} = \sum_{N=0}^\infty  \underbrace{\frac{2^{N/2}\,\alpha^N } {\sqrt {N!}\, \mathrm{e}^{ |\alpha|^2}} }_{\mbox{$ c^\alpha_N$}}\, |\phi_\mathrm{difN}\rangle_{k,k'}
\label{Eq:dif-state-cN}
\end{eqnarray}
where
\begin{eqnarray}
&\,&\hspace*{-20pt} |\phi_\mathrm{difN}\rangle_{k,k'}
\nonumber \\
&=&\! \!\frac{1}{2^{N/2}} \sum_{m=0}^N  \mathrm{e}^{\mathrm{i} (N-m) \varphi} \sqrt{\frac{N !}{m! \,(N\!-\!m)!}}    \, |m \rangle_k  \, |N\!-\!m \rangle_{k'}~~~
\nonumber \\
\label{Eq:SM-dif-N-bin-substates}
\end{eqnarray}
The states  fulfill the normalization  $\langle \Phi_\mathrm{dif} |\Phi_\mathrm{dif} \rangle= \langle \phi_\mathrm{difN} |\phi_\mathrm{difN} \rangle=1$.

\subsection{The Collective Chaotic State and its Substates}

The 2-mode collective chaotic state and its substates, describing the case in   Fig.\,\ref{Fig:2-photon-sources}\,(e), are given by the general form (\ref{Eq:SM-cha-2-mode-state-0}) which may be rewritten as,
\begin{eqnarray}
 &\,&\hspace*{-20 pt} |\Phi_\mathrm{cha}\rangle_{k,k'}
 \nonumber \\
 &=&\!\! \sum_{N=0}^\infty  \sqrt{\! \frac{\langle n \rangle^N}{\left( 1 \! +\! \langle n \rangle \right)^{N+2}}} \sum_{m=0}^N \mathrm{e}^{\mathrm{i}( \phi_m +\phi_{N-m})}  \, |m\rangle_k |N\!-\!m\rangle_{k'}~~~
 \hspace*{-15 pt}
 \nonumber \\
\label{Eq:SM-cha-2-mode-state}
\end{eqnarray}
The phase factors account for the relative phase difference between number states $|m\rangle_k$ and  $|N-m\rangle_{k'}$ in the two modes.
It also contains substates with a specific number of $N$ photons. Their probability is distributed in the form of a \emph{Bose-Einstein} distribution around the average value $2\langle n \rangle$. The state (\ref{Eq:SM-cha-2-mode-state}) may be written as,
\begin{equation}
|\Phi_\mathrm{cha}\rangle_{k,k'} =   \sum_{N=0}^\infty  \underbrace{\sqrt{\! \frac{(N+1) \langle n \rangle^N}{\left( 1 \! +\! \langle n \rangle \right)^{N+2}}} }_{\mbox{$ c^\beta_N  $}} \, |\phi_\mathrm{chaN}\rangle_{k,k'}
\label{Eq:cha-state-cN}
\end{equation}
where $\sum_{N=0}^\infty  |c^\beta_N|^2=1  $ and the substates are,
\begin{eqnarray}
\hspace*{-20pt}
|\phi_\mathrm{chaN}\rangle_{k,k'}\!\! &=& \!\! \frac{1}{\sqrt{N+1}} \sum_{m=0}^N \mathrm{e}^{\mathrm{i}\phi_{m,N-m}}  |m\rangle_k |N\!-\!m\rangle_{k'}
\label{Eq:SM-cha-N-bin-substates}
\end{eqnarray}
The states  fulfill the normalization $\langle \Phi_\mathrm{cha} |\Phi_\mathrm{cha} \rangle=\langle \phi_\mathrm{chaN} |\phi_\mathrm{chaN} \rangle=1$.

\subsection{Plots of the Substate Distributions}

\begin{figure}[!h]
\centering
\includegraphics[width=0.75\columnwidth]{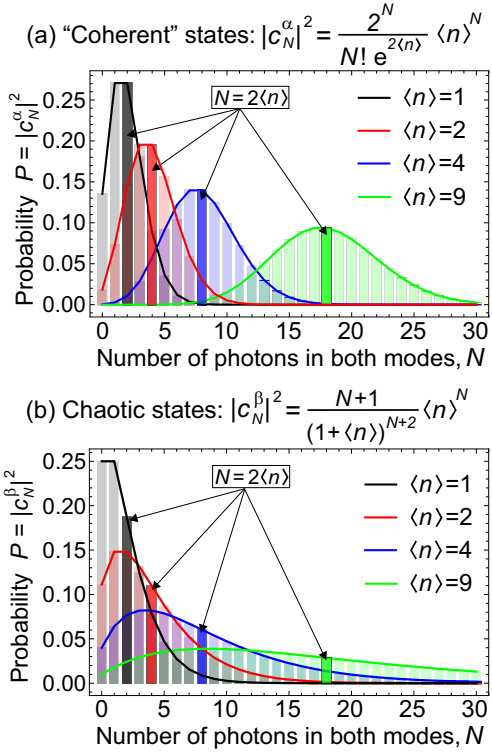}
\caption[]{Probability distributions of the substates of (a) the 2-mode collective coherent state (\ref{Eq:coh-state-cN}) and phase-diffused coherent state (\ref{Eq:dif-state-cN}), and (b) the collective chaotic state  (\ref{Eq:cha-state-cN}). For the shown cases, the collective states contain different average number of photons per mode $\langle n\rangle=1,2,4,9$, while their substates contain different total numbers of $N$ photons in both modes. The cases where $N$ reflects the distribution average $2\langle n\rangle$ are shown in enhanced colors. }
 \label{Fig:Probability-distributions}
\end{figure}

The substructure of the 2-mode collective coherent, phase-diffused coherent and chaotic states is illustrated in Fig.\,\ref{Fig:Probability-distributions} for the cases of different orders of coherence $O$, defined by an average number of photons per mode $O=\langle n\rangle=1,2,4,9$.

The cases where $2\langle n\rangle =N$ are shown in enhanced colors. The probabilities of all distributions sum to unity according to $\sum_{N=0}^\infty  |c^\alpha_N|^2=\sum_{N=0}^\infty  |c^\beta_N|^2 =1  $.

\subsection{Other Fundamental Quantum States}

\subsubsection{$N$-Photon Entangled or NOON State}

A particularly important state in quantum information science is the $N$-photon entangled state \cite{bouwmeester:2000,haroche:2001,horodecki:09,pan:2012} given by,
\begin{eqnarray}
|\phi_\mathrm{entN} \rangle_{k,k'}  =  \frac{1}{\sqrt{2}}\bigg[ |N\rangle_k |0\rangle_{k'} + \mathrm{e}^{\mathrm{i}\phi }  |0\rangle_k |N\rangle_{k'} \bigg]
 \label{Eq:SM-phi-entN}
\end{eqnarray}
which we have written in a form that reflects why it is also called a NOON state. It corresponds to $N$-photons being  emitted into the \emph{same} mode and none into the other. The specific state $N=2$ given by,
\begin{eqnarray}
|\phi_\mathrm{ent2}\rangle_{k,k'} =  \frac{1}{\sqrt{2}} \bigg[ |2\rangle_k|0\rangle_{k'} + \mathrm{e}^{ \mathrm{i}\varphi} |0\rangle_k| 2 \rangle_{k'} \bigg]
\label{Eq:SM-2-mode-ent-state}
\end{eqnarray}
describes the case in  Fig.\,2\,(b).

\subsubsection{$N$-Photon Number State}

In the complementary case, an equal number of $N/2$ photons are emitted into separate modes, described by the 2-mode number state,
\begin{eqnarray}
 |\phi_\mathrm{numN}\rangle_{k,k'} =   \left|\frac{N}{2}\right \rangle_k \left |\frac{N}{2} \right \rangle_{k'}
  \label{Eq:SM-phi-numN}
\end{eqnarray}
Owing to the indivisibility of photons the state exists only for $ N\geq 2$. The specific state with $N=2$,
\begin{eqnarray}
|\phi_\mathrm{num2}\rangle_{k,k'} =  |1\rangle_k |1\rangle_{k'}
\label{Eq:SM-2-mode-num-state}
\end{eqnarray}
describes the case in  Fig.\,2\,(c).

\subsection{Summary of Key Multi-Photon Quantum States}

The normalized 2-mode multi-photon states representing the cases with corresponding names in Fig.\,\ref{Fig:2-photon-sources} are summarized in Table\,I, for convenience.
\begin{table}[!h]
\footnotesize
\caption[] {The quantum states of light associated with Fig.\,\ref{Fig:2-photon-sources}$^\natural$}
\renewcommand{\arraystretch}{1.5}
\begin{displaymath}
\begin{array}{|l|} \hline
\mbox{\small Coherent\,states\,and\,substates:}  \\
\displaystyle |\Phi_\mathrm{coh} \rangle_{k,k'} \! = \! \sum_{N=0}^\infty   c^\alpha_N \,  |\phi_\mathrm{cohN}\rangle_{k,k'} \! = \! \sum_{N=0}^\infty    \frac{2^{N/2}\,\alpha^N } {\sqrt {N!}\, \mathrm{e}^{ |\alpha|^2}} \,  |\phi_\mathrm{cohN}\rangle_{k,k'}   \\
 \displaystyle |\phi_\mathrm{cohN}\rangle_{k,k'} \!  = \!  \frac{1}{2^{N/2}} \sum_{m =0}^N \sqrt{\frac{N !}{m! \,(N\!-\!m)!}}  ~ |m\rangle_k  \, | N\!-\!m  \rangle_{k'}      \\
\hline
\mbox{\small Phase-diffused\,coherent\,states\,and\,substates:}\\
\displaystyle |\Phi_\mathrm{dif} \rangle_{k,k'} \!= \! \sum_{N=0}^\infty   c^\alpha_N \,  |\phi_\mathrm{difN}\rangle_{k,k'} \! = \! \sum_{N=0}^\infty   \frac{2^{N/2}\,\alpha^N } {\sqrt {N!}\, \mathrm{e}^{ |\alpha|^2}} \, |\phi_\mathrm{difN} \rangle_{k,k'}  \\
 \displaystyle |\phi_\mathrm{difN} \rangle_{k,k'}\! = \!\frac{1}{2^{N/2}} \sum_{m=0}^N  \mathrm{e}^{\mathrm{i} (N-m) \varphi} \sqrt{\frac{N !}{m! \,(N\!-\!m)!}}   \, |m \rangle_k  \, |N\!-\!m \rangle_{k'} \\
\hline
\mbox{\small Chaotic\,states\,and\,substates:} \\
\displaystyle |\Phi_\mathrm{cha}\rangle_{k,k'} \!= \!\! \sum_{N=0}^\infty \! c^\beta_N  |\phi_\mathrm{chaN}\rangle_{k,k'}\!= \!\!   \sum_{N=0}^\infty \!\sqrt{\! \frac{(N+1)\langle n \rangle^N}{\left( 1 \! +\! \langle n \rangle \right)^{N+2}}} \, |\phi_\mathrm{chaN}\rangle_{k,k'}  \\
\displaystyle  |\phi_\mathrm{chaN}\rangle_{k,k'} \!=\! \frac{1}{\sqrt{N+1}} \sum_{m=0}^N \mathrm{e}^{\mathrm{i}\phi_{m,N-m}}\,  |m\rangle_k |N\!-\!m\rangle_{k'}  \\
\hline
\mbox{\small $N$-photon\,entangled\,(NOON) state:}\\
\displaystyle  |\phi_\mathrm{entN}\rangle_{k,k'} =  \frac{1}{\sqrt{2}} \left(  |N\rangle_k|0\rangle_{k'} +\mathrm{e}^{\mathrm{i}\varphi} |0\rangle_k| N \rangle_{k'}    \right)\\
\hline
\mbox{\small $N$-photon\,number\,state ($N\!>\!1$):}\\
\displaystyle |\phi_\mathrm{numN}\rangle_{k,k'} =  |N/2\rangle_k |N/2\rangle_{k'} \\
\hline
\end{array}
\end{displaymath}
$^\natural$
The collective states  $|\Phi_i\rangle_{k,k'}$ have $\langle n\rangle$ photons in each mode, and for the coherent states we have $|\alpha|^2=\langle n\rangle$. The substates $|\phi_j\rangle_{k,k'}$ have $N$ photons in both modes.
All states have unit normalization and $\sum_{N=0}^\infty |c^\alpha_N|^2 \!=\!\sum_{N=0}^\infty  |c^\beta_N|^2\!=\!1$.
\newline
We also have the important sum rules in first and second order:\\
$ \sum_{N=0}^\infty \frac{N}{2} \, |c^\alpha_N|^2 = \sum_{N=0}^\infty \frac{N}{2} \, |c^\beta_N|^2  =\langle n \rangle$  and \\
$ \sum_{N=0}^\infty \frac{N(N-1)}{4} \, |c^\alpha_N|^2 =  \sum_{N=0}^\infty \frac{N(N-1)}{6} \, |c^\beta_N|^2=   \langle n \rangle^2$.
\label{T:quantum-states}
\end{table}

\section{First Order Diffraction Patterns}

The key part of the calculation of the first order patterns $\left \langle   \mathbf  P^{(1)}(\vec \rho_1,   \vec \rho_2  )  \right\rangle$ is the calculation of the matrix elements or expectation values of the operators $\mathbf  X$ and $\mathbf  Y$ in (\ref{Eq:SM-term-X}) and (\ref{Eq:SM-term-Y}) with the quantum states in Table\,I. The lengthy calculations of the matrix elements involve application of the well known quantum mechanical rules for the action of creation and annihilation operators on the number substates \cite{Dirac-book,scully-zubairy} and utilize sum rules like those listed below Table I. In this section we present the derivation of the diffraction pattern for the example of the coherent state and its substates. The similar derivations of the patterns for the other states in Table\,I are given in Appendix A.

\subsection{Coherent State}

For the \emph{collective coherent state} (\ref{Eq:coh-state-cN}) the non-vanishing matrix elements of the terms in (\ref{Eq:SM-term-X}) and
(\ref{Eq:SM-term-Y}) are given by,
\begin{eqnarray}
&\,& \langle \Phi_\mathrm{coh} |\mathbf a^\dag_{\vec{k} } \, \mathbf   a_{ \vec{k} } |\Phi_\mathrm{coh}\rangle
\! = \! \langle \Phi_\mathrm{coh} |\mathbf a^\dag_{\vec{k}' } \, \mathbf   a_{ \vec{k}' } |\Phi_\mathrm{coh}\rangle
 \nonumber \\
 &\,&  = \langle \Phi_\mathrm{coh} |\mathbf a^\dag_{\vec{k} } \, \mathbf   a_{ \vec{k}' } |\Phi_\mathrm{coh}\rangle
\! = \! \langle \Phi_\mathrm{coh} |\mathbf a^\dag_{\vec{k}' } \, \mathbf   a_{ \vec{k}  } |\Phi_\mathrm{coh}\rangle
\nonumber \\
 &\,&=   \sum_{N=0}^\infty \frac{ |\alpha|^{2N}}{\mathrm{e}^{ 2 |\alpha|^2}} \,  \sum_{m=0}^N  \,   \frac{m}{m!(N-m)!  }  = |\alpha|^{2}
\label{Eq:SM-matrix-elem-coh-state}
\end{eqnarray}
The 1-photon detection probability is obtained as,
\begin{eqnarray}
&\,& \hspace*{-30pt}\left \langle   \mathbf  P^{(1)}(\vec x_1, \vec x_2  )  \right\rangle_\mathrm{\!coh} =\left \langle \Phi_\mathrm{coh} \left | \mathbf  P^{(1)}(\vec x_1, \vec x_2  ) \right | \Phi_\mathrm{coh}  \right\rangle
\nonumber \\
&\,& =
\frac{1}{2}\bigg \{
\underbrace{  |\alpha|^2 \mathrm{e}^{  \mathrm {i} \vec{k}  \cdot (\vec{x}_2 - \vec{x}_1) }
+ |\alpha |^2 \mathrm{e}^{  \mathrm {i} \vec{k}' \cdot (\vec{x}_2 - \vec{x}_1) }}_{\mbox{$\langle \Phi_\mathrm{coh} |\mathbf X|\Phi_\mathrm{coh}\rangle $}}
\nonumber \\
&\,& ~ ~
+ \underbrace{|\alpha|^2 \, \mathrm{e}^{  \mathrm {i} (\vec{k}' \cdot \vec{x}_2  - \vec{k}  \cdot \vec{x}_1)  }
  +   |\alpha|^2 \,  \mathrm{e}^{  \mathrm {i}  (\vec{k}  \cdot \vec{x}_2 - \vec{k}' \cdot \vec{x}_1) }}_{\mbox{$\langle \Phi_\mathrm{coh} |\mathbf Y|\Phi_\mathrm{coh}\rangle$}} \bigg \}
\label{Eq:SM-prob-1-photon-coh-state}
\end{eqnarray}

We can now express the coordinates $(\vec k, \vec x)$  in terms of $(\vec r,\vec \rho)$ in Fig.\,1\,(a) and by assuming that the detector plane is at a large distance $z_0$ from the source we have the relations \cite{stohr-AOP},
\begin{eqnarray}
\vec k_\mathrm{X} \! \cdot \! \vec x_j \simeq k  z_0 - \frac{k}{z_0} \vec r_X \! \cdot \!\vec \rho_j,~~~
\end{eqnarray}
where $X=A,B$ and $\vec k_\mathrm{A}=\vec k $ and $\vec k_\mathrm{B}=\vec k'$ and $j=1,2$. For our assumed geometry in Fig.\,1\,(b) with point sources located at $\vec r_A=-\vec r_B$, separated by $\ell = | \vec r_A-\vec r_B|$, and $ \vec \rho_2 \parallel \vec \rho_1 $ we then obtain,
\begin{eqnarray}
\hspace*{-10pt}\left \langle   \mathbf  P^{(1)}(\vec \rho_1, \vec \rho_2 )  \right\rangle_\mathrm{coh}\! = \! 2|\alpha|^2 \cos\left[\frac{ k \ell}{2 z_0}  \rho_1  \right]\cos\left[ \frac{ k \ell}{2 z_0}  \rho_2  \right]
\label{Eq:SM-P1-coh-point-sorces}
\end{eqnarray}

The case of \emph{finite-size} double-slit sources of widths $a$ is evaluated by integrating over all points in the slits.  However, the integration needs to be performed over the probability \emph{amplitudes} (\ref{Eq:Glauber-1-P-amplitudes}). Only then is the detection probability calculated as the absolute value squared of the integrated amplitudes. This yields an additional envelope function and with $|\alpha|^2=\langle n\rangle$ we obtain,
\begin{eqnarray}
\hspace*{-10pt} \left \langle   \mathbf  P^{(1)}(\vec \rho_1,   \vec \rho_2  )  \right\rangle_\mathrm{\!coh}
\! &=& \! 2 \langle n\rangle \cos\! \left[ \frac{k\, \ell}{2 z_0}  \rho_1  \right] \cos\! \left[ \frac{k\, \ell}{2 z_0}  \rho_2  \right]
\nonumber \\
&\,&  \times \, \mathrm{sinc} \! \left[  \frac{k\, a}{ 2 z_0} \rho_1 \right] \mathrm{sinc} \! \left[  \frac{k\, a}{ 2 z_0} \rho_2 \right]
\label{Eq:SM-1-photon-double-slit-coh}
\end{eqnarray}

\subsection{$N$-Photon Substate of Coherent State}

The diffraction pattern of the 2-mode \emph{coherent substate} (\ref{Eq:SM-coh-N-bin-substates}) is calculated by use of the matrix elements,
\begin{eqnarray}
 &\,&\langle \phi_\mathrm{cohN}|  \mathbf a^\dag_{\vec{k} } \, \mathbf a_{\vec{k}' }   |\phi_\mathrm{cohN}\rangle  = \langle \phi_\mathrm{cohN}|  \mathbf a^\dag_{\vec{k}' } \, \mathbf a_{\vec{k}}   |\phi_\mathrm{cohN}\rangle
 \nonumber \\
 &\,& ~~ =
\langle \phi_\mathrm{cohN}|  \mathbf a^\dag_{\vec{k}' } \, \mathbf   a_{\vec{k}'} |\phi_\mathrm{cohN}\rangle =\langle \phi_\mathrm{cohN}|  \mathbf a^\dag_{\vec{k}  } \, \mathbf   a_{\vec{k}} |\phi_\mathrm{cohN}\rangle
 \nonumber \\
 &\,& ~~ = \frac{1}{2^N } \sum_{n =0}^N  \frac{n\, N !}{n! \,(N\!-\!n)!} =\frac{N}{2}
\end{eqnarray}
The diffraction pattern $\left \langle \phi_\mathrm{cohN}| \mathbf  P^{(1)}(\vec \rho_1, \vec \rho_2  )|\phi_\mathrm{cohN}\right \rangle$  is hence the collective coherent state result (\ref{Eq:SM-1-photon-double-slit-coh}) with $2 \langle n\rangle$ replaced by $N$, i.e.
\begin{eqnarray}
\hspace*{-10pt} \left \langle   \mathbf  P^{(1)}(\vec \rho_1,   \vec \rho_2  )  \right\rangle_\mathrm{\!cohN}
\! &=& \! N \cos\! \left[ \frac{k\, \ell}{2 z_0}  \rho_1  \right] \cos\! \left[ \frac{k\, \ell}{2 z_0}  \rho_2  \right]
\nonumber \\
&\,&  \times \, \mathrm{sinc} \! \left[  \frac{k\, a}{ 2 z_0} \rho_1 \right] \mathrm{sinc} \! \left[  \frac{k\, a}{ 2 z_0} \rho_2 \right]
\label{Eq:SM-1-photon-double-slit-cohN}
\end{eqnarray}

 When the pattern is summed over $N$ with the proper weight coefficients, we see from the relation
\begin{eqnarray}
\sum_{N=0}^\infty N \, |c^\alpha_N|^2 \!=\! 2\langle n\rangle
\end{eqnarray}
that the coherent result  (\ref{Eq:SM-1-photon-double-slit-coh}) is obtained.

\subsection{Plots of the First Order Patterns}

\begin{figure}[!h]
\centering
\includegraphics[width=1.0\columnwidth]{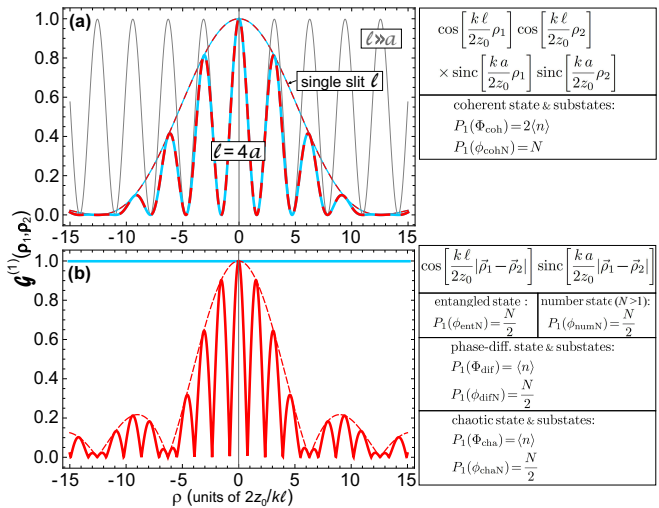}
\caption[]{ First order double slit diffraction patterns for the quantum states of light in Table\,I, with the blue and red colors representing the two detection schemes  in Fig.\,\ref{Fig:geometry}\,(b). The shape of the patterns $\boldsymbol {{\cal G}}^{(1)}  (\vec \rho_1, \vec \rho_2  )$ and scaling factors $P_1$ are given on the right. (a) Patterns $\boldsymbol {{\cal G}}^{(1)}  (\vec \rho_1, \vec \rho_2  )$ for the coherent state and its substates representing the cases  in Fig.\,\ref{Fig:2-photon-sources}\,(a). The gray curve assumes point like slits of width $a \ll \ell$, the thick red-blue pattern is that for $a=\ell/4$, and the thin red-blue line represents the sinc$^2$ envelope function which remains if the two slits are joined into a single slit of length $\ell$.  (b) Patterns for all other quantum states and cases in Fig.\,\ref{Fig:2-photon-sources}\,(b)--(e).
}
\label{Fig:1st-ord-patterns}
\end{figure}

The calculated shapes $\boldsymbol {{\cal G}}^{(1)}  (\vec \rho_1, \vec \rho_2  )$  of the \emph{first order} patterns defined in (\ref{Eq:Det-prob-O}) for the states in Table\,I  are plotted in Fig.\,\ref{Fig:1st-ord-patterns}. The shapes and  scaling factors $P_1$ for the states are given for convenience on the right.  Remarkably, the different quantum states result in only two kinds of characteristic patterns, revealing a degeneracy of the patterns for several of the states. The patterns are identical to those in wave and statistic optics for the limiting cases of ``coherent'' and ``incoherent'' light \cite{stohr-AOP,Goodman-SO}.

The  \emph{coherent states} and their substates produce the same characteristic pattern, shown as dashed thick blue and red curves in Fig.\,\ref{Fig:1st-ord-patterns}\,(a) for $a=\ell/4$. The patterns are the same for the blue and red detection geometries in Fig.\,\ref{Fig:geometry}\,(b). This is due to the fact that the detection probability $ \left \langle  \mathbf  P^{(1)}(\vec \rho_1, \vec \rho_2  ) \right \rangle$ \emph{factors} in the coordinates $\vec \rho_1$ and $\vec \rho_2$  \cite{glauber:65}. For convenience we also show the two limits of point-like slits of width $a \ll \ell$ as a gray curve,  and a single slit of length $\ell$, represented by the thin red-blue sinc$^2$ envelope function.

All other states in Table\,I  form the patterns shown in Fig.\,\ref{Fig:1st-ord-patterns}\,(b). The detection probability does not factor in the coordinates $\vec \rho_1$ and $\vec \rho_2$ and the source is ``incoherent'' in first order. This results in \emph{constant} photon counts for the blue detection geometry in Fig.\,\ref{Fig:geometry}\,(b) with a diffraction structure observed only for the red detection geometry.

In all cases, the \emph{shapes} of the patterns $\boldsymbol {{\cal G}}^{(1)}  (\vec \rho_1, \vec \rho_2  )$ in Fig.\,\ref{Fig:1st-ord-patterns} are the same for the three \emph{collective states} containing an average of $2\langle n\rangle$ photons  and their respective \emph{substates} containing $N$ photons. The shapes are even independent of the number of photons, $N$, contained in the substates. Emission of more than single photons, $N\!>\!1$, only increases the scaling factor $P_1$, i.e. the overall detection probability. This explains the experimental fact that the first order pattern for the case shown in Fig.\,\ref{Fig:2-photon-sources}\,(a) is independent of whether the slits are illuminated by a thermal source that has been made first order coherent by energy (monochromator) and spatial (aperture) filtering or by a laser, although the light is coherent to different orders in QED for the two cases.

This means that the nature of the collective states remains preserved in the coefficients and phases of their respective substates, independent of the number of photons, $N$, they contain. The patterns in  Fig.\,\ref{Fig:1st-ord-patterns}  may therefore be recorded with 1-photon detectors that pick out the $N=1$ state or charge integrating detectors which detect arbitrary numbers of arriving photons. Note that even the  collective states with $\langle n\rangle=1$ contain different number of photons $N=1,2,3...$ as shown in Fig.\,\ref{Fig:Probability-distributions}. By use of charge or photon-number integrating detectors the diffraction patterns therefore accumulate through the arrival of photons whose number statistically varies between $N=1$ and about $N=6$.

This shows that the first order patterns do not depend on the  \emph{photon degeneracy parameter} or number of emitted photons per mode, $\langle n\rangle$ or $N/2$. This agrees with Dirac's statement that photons do not interfere with each other \cite{Dirac-book}. The subtlety of this statement, however, is revealed by the fact that photons in the same mode are in principle \emph{indistinguishable} as emphasized by Ou \emph{et al.} \cite{Ou:2007,Ou:2017}, and one would therefore expect them to interfere. The dilemma is resolved by the fact that a meaningful diffraction pattern always corresponds to the accumulation of a large number of detection events. Any interference structure that may be present when only few photons arrive in coincidence are increasingly averaged out upon appearance of a statistically meaningful pattern \cite{javanainen:96,Mandel:99}.

In x-ray science, the \emph{independence} of the Bragg diffraction pattern on the degeneracy parameter has the important consequence that all diffraction patterns observed for the last 100 years with weak sources such as R\"{o}ntgen tubes remain the same when recorded in a single shot with an x-ray free electron laser, despite an increase of the degeneracy parameter by about 25 orders of magnitude \cite{stohr-AOP}.
This means that it does not matter whether the pattern assembles one photon at a time or in a ``single shot''. With increasing degeneracy parameter the statistics of the pattern is simply improved. The pattern may be recorded much faster, even in a single few-femtosecond shot \cite{wang:12}. Hence no new x-ray diffraction theory is needed to describe first order patterns.

\subsection{Degree of First Order Coherence}

The first order diffraction probabilities $\left \langle \mathbf  P^{(1)}(\vec \rho_1, \vec \rho_2  )  \right\rangle$ for the cases in Fig.\,\ref{Fig:1st-ord-patterns}, are readily converted by means of (\ref{Eq:SM-Degree-of-coherence}) into the corresponding degrees of first order coherence. For the double slit geometry we have
\begin{eqnarray}
g^{(1)}(\vec \rho,-\vec \rho) =\frac{\left \langle \mathbf  P^{(1)}(\vec \rho, -\vec \rho  )  \right\rangle}
{ \left \langle \mathbf  P^{(1)}(\vec \rho, \vec \rho  )  \right\rangle  }
\label{Eq:First-order-degree-of-coherence}
\end{eqnarray}
For the coherent case in Fig.\,\ref{Fig:1st-ord-patterns}\,(a) we obtain with (\ref{Eq:SM-1-photon-double-slit-coh}) the expected result,
\begin{eqnarray}
g^{(1)} (\vec \rho ,-\vec \rho)= 1
\label{Eq:g1-coh}
\end{eqnarray}
The ``incoherent'' cases in Fig.\,\ref{Fig:1st-ord-patterns}\,(b) all yield the same expression
\begin{eqnarray}
g^{(1)} (\vec \rho,-\vec \rho )=  \cos \! \left[ \frac{k\, \ell}{  z_0}  \rho   \right]  \mathrm{sinc}  \! \left[  \frac{k\, a}{ z_0} \rho \right]
\label{g1-incoherent-states}
\end{eqnarray}
Hence we have $g^{(1)} (\vec \rho,-\vec \rho )= \boldsymbol {{\cal G}}^{(1)}  (\vec \rho , - \vec \rho )$ for these cases, represented by red pattern Fig.\,\ref{Fig:1st-ord-patterns}\,(b).

\subsection{Reduction of the First-Order Quantum to the Wave Formalism}

Feynman's and Glauber's \emph{first order} quantum formulations of diffraction and the classical or statistical optics \cite{Born-Wolf,Goodman-SO} descriptions give the same diffraction patterns. The formulations are all based on propagation paths from source to detection points defined by geometric trajectories as illustrated in  Fig.\,\ref{Fig:geometry}, and the diffraction patterns correspond to the interference of \emph{probability amplitudes}.

In \emph{Glauber's quantum optics formulation}, used here, the probability amplitudes (\ref{Eq:Glauber-1-P-amplitudes}) contain \emph{no birth phases}. Photon birth and destruction are treated by Hermitian conjugate operators that create photons out of the quantum vacuum and destroy them back into it. Different diffraction patterns arise when the expectation value of a detection probability operator  $\mathbf  P^{(1)}(\vec \rho_1, \vec \rho_2  )$ is calculated by use of different quantum states created in the source.

In \emph{Feynman's formulation}, the probability amplitudes (\ref{Eq:Feynman-1-P-amplitudes}) have the same structure as Glauber's amplitudes (\ref{Eq:Glauber-1-P-amplitudes}), but the creation operators  $\mathbf  a^\dagger_{\vec{k}}$ are replaced by birth phase factors  $\mathrm{e}^{ - \mathrm {i} \alpha_{\vec{k}}}$, and the destruction process is treated as the complex conjugate of the birth process. The tricky part in Feynman's formulation is the evaluation of the quantum mechanical expectation value of the probability operator, which requires an \emph{average over birth phases} for different cases. While the description of the extreme coherent and chaotic cases is simple since the birth phases are either the same or random, the treatment of the intermediate cases of partial coherence are non-trivial. This contrasts  Glauber's formulation where the statistical average is given by the expectation value of $\mathbf  P^{(1)}(\vec \rho_1, \vec \rho_2  )$ for a given quantum state of light.

The conventional \emph{wave formalism} of diffraction emerges from the quantum descriptions by approximating  the average over all photon emission directions by the concept of a spherical wave. One then makes the \emph{ad hoc} assumption that all waves are created with the same birth phase. The distinction between coherence and incoherence arises from geometry alone. Spherical waves emitted from  different source points are said to be coherent at a distant observation point, if one may approximate them by \emph{plane waves} within a finite solid angle ``observation cone'' extending backwards from  the observation point to the source points. The picture may also be turned around by defining a coherent \emph{emission  cone} rather than a coherent \emph{observation  cone}. This simple geometrical concept means that within the solid angle coherence cone the curvature of the spherical wave is negligible on the scale of the wavelength. Classically one simply calculates the ``intensity'' at different points in the observation plane as the absolute value squared of interfering wavefields. The measurement process of the intensity remains unspecified.

One might have expected that an incoherent or chaotic source, reflected by the \emph{chaotic state} in quantum optics,  does not give rise to a diffraction pattern at all. The reason for its existence, revealed by the red pattern in Fig.\,\ref{Fig:1st-ord-patterns}\,(b), lies in the fact that for the spatial phenomenon of diffraction one assumes space-time separability and temporal coherence, i.e. that the photons have the same well defined energy or wavelength $\lambda$. If the source were chaotic in both space and time there would indeed be no position dependent diffraction structure.

The pattern in Fig.\,\ref{Fig:1st-ord-patterns}\,(b) for chaotic quantum states produced in the source, follows in wave or statistical optics from the powerful van Cittert--Zernike theorem \cite{vancittert:34,zernike:38,Goodman-SO,wolf:2007,stohr-AOP}. It picks out the coherent fraction of light emitted by a source, even if the source is chaotic. In classical electromagnetism, radiation (i.e. acceleration fields \cite{jackson}) can only separate from the charge if it is defined at least over the dimension of its average wavelength. Hence any light-emitting source contains a coherent fraction that arises from waves emitted from the minimum coherence area of order $\lambda^2$ \cite{beran-parrent,deutsch-garrison:91}. In classical optics, the 1D pattern in Fig.\,\ref{Fig:1st-ord-patterns}\,(b) arises from waves emitted from regions of lateral size $\simeq \lambda$ within the two slits that interfere at detection points.

The more fundamental photon nature of light, described by QED, just happens to be describable in first order by the conventional wave formalism. One has to realize, however, that the classical theory is based on \emph{ad hoc} assumptions or postulates which make it work. The fundamental origin of these assumptions, like the perceived existence of spherical waves and the validity of the Fresnel-Huygens principle, emerges in lowest order QED as a consequence of the interference of single photon probability amplitudes associated with all possible photon paths to a given detection point.

The independence of the first order quantum pattern of the photon degeneracy parameter also reveals why the detection process did not have to be specified in the classical formulation. The  diffracted ``intensity'' is simply calculated as the absolute value squared of the ``wave field'' and the quality or statistics of the pattern improves linearly with ``intensity''. Historically, this allowed the use of wave concept long before the true photon nature of light was known.

In the following sections we will show that only QED can account for more sophisticated diffraction experiments carried out by changing the detection process, revealing the intrinsic limitation of the wave theory of light.

\section{Second Order Diffraction Patterns}

The \emph{second order} diffraction patterns $\left \langle   \mathbf  P^{(2)}(\vec \rho_1,   \vec \rho_2  )  \right\rangle$ for the quantum states in Table\,I are calculated by use of the  matrix elements of the  four second order terms (\ref{Eq:SM-term-A})--(\ref{Eq:SM-term-D}) with the quantum states. As for the first order case we limit the full derivation of the diffraction patterns to the example of the coherent state and its substates. The similar derivations for the other states are given in Appendix B.

\subsection{Coherent State}

For the collective coherent state  (\ref{Eq:coh-state-cN}), the relevant matrix elements for term $\langle \mathbf A \rangle $ are given by,
\begin{eqnarray}
 &\,& \hspace*{-30pt}\langle \Phi_\mathrm{coh} |  \mathbf a_{\vec{k}}^{\dagger} \,\mathbf  a_{ \vec{k}'}^{\dagger} \, \mathbf a_{\vec{k} }  \,   \mathbf  a_{ \vec{k}'} |\Phi_\mathrm{coh}\rangle = \langle \Phi_\mathrm{coh} |  \mathbf a_{\vec{k}'}^{\dagger} \,\mathbf  a_{ \vec{k}}^{\dagger} \, \mathbf a_{\vec{k}' }  \,   \mathbf  a_{ \vec{k} } |\Phi_\mathrm{coh}\rangle
 \nonumber \\ &\,& = \mathrm{e}^{- 2 |\alpha|^2}\sum_{N=0}^\infty  |\alpha|^{2N} \, \sum_{n=0}^N \frac{n(N\!-\!n)  }{ n!\, (N-n)! }
= |\alpha|^4
\label{Eq:SM-term-A-coh}
\end{eqnarray}
For term  $\langle \mathbf B\rangle $ we have
\begin{eqnarray}
 &\,& \hspace*{-30pt}\langle \Phi_\mathrm{coh} |  \mathbf a_{\vec{k}}^{\dagger} \,\mathbf  a_{ \vec{k} }^{\dagger} \, \mathbf a_{\vec{k} }  \,   \mathbf  a_{ \vec{k}} |\Phi_\mathrm{coh}\rangle =\langle \Phi_\mathrm{coh} |  \mathbf a_{\vec{k}'}^{\dagger} \,\mathbf  a_{ \vec{k} '}^{\dagger} \, \mathbf a_{\vec{k}' }  \,   \mathbf  a_{ \vec{k}'} |\Phi_\mathrm{coh}\rangle
 \nonumber \\
&=& \mathrm{e}^{- 2 |\alpha|^2}\sum_{N=0}^\infty  |\alpha|^{2N} \, \sum_{n=0}^N \frac{n(n-1)  }{ n!\, (N-n)! }
=|\alpha|^4
\label{Eq:SM-term-B-coh-1}
\end{eqnarray}
and
\begin{eqnarray}
\langle \Phi_\mathrm{coh} |  \mathbf a_{\vec{k}}^{\dagger} \,\mathbf  a_{ \vec{k} }^{\dagger} \, \mathbf a_{\vec{k}' }  \,   \mathbf  a_{ \vec{k}'} |\Phi_\mathrm{coh}\rangle  &=&\langle \Phi_\mathrm{coh} |  \mathbf a_{\vec{k}'}^{\dagger} \,\mathbf  a_{ \vec{k}' }^{\dagger} \, \mathbf a_{\vec{k} }  \,   \mathbf  a_{ \vec{k} } |\Phi_\mathrm{coh}\rangle
\nonumber \\
&=& |\alpha|^4
\label{Eq:SM-term-B-coh-2}
\end{eqnarray}

Similarly we obtain the matrix elements for terms $\langle \mathbf C \rangle $ and $\langle \mathbf D \rangle $
\begin{eqnarray}
 \langle \Phi_\mathrm{coh} |  \mathbf a_{\vec{k}}^{\dagger} \,\mathbf  a_{ \vec{k} }^{\dagger} \, \mathbf a_{\vec{k}  }  \,   \mathbf  a_{ \vec{k}'} |\Phi_\mathrm{coh}\rangle
 &=&
 \langle \Phi_\mathrm{coh} |  \mathbf a_{\vec{k}}^{\dagger} \,\mathbf  a_{ \vec{k}' }^{\dagger} \, \mathbf a_{\vec{k} ' }  \,   \mathbf  a_{ \vec{k}'} |\Phi_\mathrm{coh}\rangle
 \nonumber \\
\langle \Phi_\mathrm{coh} |  \mathbf a_{\vec{k}'}^{\dagger} \,\mathbf  a_{ \vec{k} }^{\dagger} \, \mathbf a_{\vec{k} }  \,   \mathbf  a_{ \vec{k}} |\Phi_\mathrm{coh}\rangle
&=&
\langle \Phi_\mathrm{coh} |  \mathbf a_{\vec{k}'}^{\dagger} \,\mathbf  a_{ \vec{k}' }^{\dagger} \, \mathbf a_{\vec{k} ' }  \,   \mathbf  a_{ \vec{k} } |\Phi_\mathrm{coh}\rangle
 \nonumber \\
&=& |\alpha|^4
\end{eqnarray}

The second order detection probability becomes,
\begin{eqnarray}
&\,&\hspace*{-15pt}   \left \langle \mathbf  P^{(2)}(\vec x_1, \vec x_2  )  \right\rangle_\mathrm{\!coh}
\!= \!
\underbrace{  \frac{ |\alpha|^4}{2}  \!+ \!\frac{ |\alpha|^4}{2}\cos[(\vec{k}  -\vec{k}') \cdot (\vec{x}_1 - \vec{x}_2)]}_{ \mbox{$\langle \Phi_\mathrm{coh} |\mathbf  A| \Phi_\mathrm{coh}\rangle/4$ }}
\nonumber \\
&\,&  \!  +\!
\underbrace{ \frac{ |\alpha|^4}{2}\!+ \! \frac{ |\alpha|^4}{4}   \,
\mathrm{e}^{ - \mathrm{i} (\vec{k} -\vec{k}')\cdot (\vec{x}_1 +\vec{x}_2) }
\!+\!
 \frac{ |\alpha|^4}{4} \,
\mathrm{e}^{  \mathrm{i}(\vec{k} -\vec{k}') \cdot (\vec{x}_1 + \vec{x}_2) } }_{ \mbox{$\langle \Phi_\mathrm{coh} |\mathbf  B| \Phi_\mathrm{coh}\rangle/4$ }}
\nonumber \\
&\,&  \!+ \! \underbrace{  \frac{ |\alpha|^4}{2}\,\mathrm{e}^{ - \mathrm{i} (\vec{k} -\vec{k}')\cdot \vec{x}_1 }
\!+\!
 \frac{ |\alpha|^4}{2} \, \mathrm{e}^{\mathrm{i} (\vec{k} -\vec{k}') \cdot \vec{x}_1 } }_{ \mbox{$\langle \Phi_\mathrm{coh} |   \mathbf C| \Phi_\mathrm{coh}\rangle/4$ }}
\nonumber \\
&\,& \!+ \!  \underbrace{   \frac{ |\alpha|^4}{2} \, \mathrm{e}^{ - \mathrm{i} (\vec{k} -\vec{k}')\cdot \vec{x}_2 }
\!+\!
  \frac{ |\alpha|^4}{2}  \, \mathrm{e}^{\mathrm{i} (\vec{k} -\vec{k}') \cdot \vec{x}_2 }}_{ \mbox{$\langle \Phi_\mathrm{coh} |\mathbf  D| \Phi_\mathrm{coh}\rangle/4$}}
\label{Eq:SM-prob-coh-state-expression}
\end{eqnarray}
where we have identified the origin of the four terms by underbrackets. By use of $|\alpha|^2\!=\!\langle n\rangle$ and the identities $1+\cos[x]=2\cos^2[x/2]$ and  $\cos a + \cos b \!=\!2 \cos \left[\frac{ a-b }{2}\right]   \cos \left[\frac{ a+b }{2}\right]$ this becomes,
\begin{eqnarray}
&\,& \hspace*{-15pt}
\left \langle \mathbf  P^{(2)}(\vec x_1, \vec x_2  )  \right\rangle_\mathrm{\!coh}
= \langle n \rangle^2 \bigg\{ \cos^2 \! \left [ \frac{1}{2}(\vec{k} \! - \! \vec{k}') \! \cdot \! (\vec {x}_1 \! - \! \vec x_2)\right ]
 \nonumber \\   &\,&
\hspace*{-10pt}+    \cos^2 \! \left [ \frac{1}{2} (\vec{k} \! - \! \vec{k}') \! \cdot \!(\vec {x}_1 \!+ \!\vec x_2)\right ]
 \nonumber \\   &\,&
 \hspace*{-10pt}  +      2  \cos \!\left [  \frac{1}{2}(\vec{k} \! - \! \vec{k}') \! \cdot \!(\vec {x}_1 \! - \! \vec x_2)\right ]
\cos \!  \left [ \frac{1}{2} (\vec{k} \! - \! \vec{k}') \! \cdot \! (\vec {x}_1 \!+ \!\vec x_2)\right ]\bigg\}
\hspace*{-15pt}
\nonumber \\
 \label{Eq:SM-phi-0-coh-pattern}
\end{eqnarray}

The pattern is seen to become symmetrical and  the last term is just the interference term  of the amplitudes associated with the first two terms according to,
\begin{eqnarray}
\langle \mathbf  C \rangle  + \langle \mathbf D\rangle = 2 \sqrt{\langle \mathbf  A \rangle \langle \mathbf  B \rangle}
\label{Eq:SM-binomial-interference-term}
\end{eqnarray}

Converting to the $(\vec r,\vec \rho)$ coordinates, we obtain for the slit separation $\ell\!=\!|\vec r_\mathrm{A} -\vec r_\mathrm{B}|$ and detector positions $ \vec \rho_2 \parallel \vec \rho_1 $
\begin{eqnarray}
\left \langle \mathbf  P^{(2)}(\vec \rho_1, \vec \rho_2  )  \right\rangle_\mathrm{\!coh} \!=\! 4 \langle n \rangle^2 \cos^2 \!\left[\frac{k\, \ell}{2 z_0}  \rho_1 \right] \cos^2 \! \left[\frac{k\, \ell}{2 z_0}   \rho_2 \right]~~~
\label{Eq:SM-2-photon-coh-2-points}
\end{eqnarray}

For the case of two slits of width $a$, we first integrate the probability \emph{amplitude} (\ref{Eq:G2-Glauber-wavefunction}) over the points in the slits, and then calculate the detection probability as the absolute value squared of the integrated amplitudes. This changes (\ref{Eq:SM-2-photon-coh-2-points}) to,
\begin{eqnarray}
\left \langle \mathbf  P^{(2)}(\vec \rho_1, \vec \rho_2  )  \right\rangle_\mathrm{\!coh} \!\! &=& 4 \langle n \rangle^2 \cos^2 \!\left[\frac{k\, \ell}{2 z_0}  \rho_1 \right] \cos^2 \! \left[\frac{k\, \ell}{2 z_0}   \rho_2 \right]
\nonumber \\
&\,& \hspace*{-5 pt}\times
 \mathrm{sinc}^2 \!\left[  \frac{k\, a}{2 z_0}  \rho_1 \right]\mathrm{sinc}^2 \!\left[  \frac{k\, a}{2 z_0}  \rho_2 \right]
 \label{Eq:SM-2-photon-coh-2-slit}
\end{eqnarray}
The detection probability now factors into separate symmetric contributions from points  $\vec \rho_1$ and $\vec \rho_2$, and the same pattern is observed for both detection schemes.

\subsection{$N$-Photon Substate of Coherent State}

Similarly, the  coherent substate  (\ref{Eq:SM-coh-N-bin-substates}) gives equal contributions from all terms $\mathbf A-\mathbf D$ expressed by the matrix elements,
\begin{eqnarray}
&\,& \langle \phi_\mathrm{cohN}|  \mathbf a^\dag_{\vec{k} } \, \mathbf   a^\dag_{ \vec{k}'} \mathbf a_{\vec{k}  } \, \mathbf   a_{ \vec{k}'}  |\phi_\mathrm{cohN}\rangle
\nonumber \\ &\,& =
\langle \phi_\mathrm{cohN}|  \mathbf a^\dag_{\vec{k} } \, \mathbf   a^\dag_{ \vec{k} } \mathbf a_{\vec{k}  } \, \mathbf   a_{ \vec{k}}  |\phi_\mathrm{cohN}\rangle
\nonumber \\ &\,& =
\langle \phi_\mathrm{cohN}|\mathbf a^\dag_{\vec{k} } \, \mathbf   a^\dag_{ \vec{k} } \mathbf a_{\vec{k}' } \, \mathbf   a_{ \vec{k}'} |\phi_\mathrm{cohN}\rangle
\nonumber \\ &\,& =
\langle \phi_\mathrm{cohN}|\mathbf a_{\vec{k} }^{\dagger} \, \mathbf a_{\vec{k} }^{\dagger}\, \mathbf a_{\vec{k} } \, \mathbf  a_{ \vec{k}'}|\phi_\mathrm{cohN}\rangle
\nonumber \\
&\,&  =\frac{1}{2^{N }} \sum_{n =0}^N n(n-1) \, \frac{N !}{n! \,(N\!-\!n)!}  =\frac{1}{4} N(N-1)~~~
\end{eqnarray}
and the same when exchanging $k$ and $k'$. The pattern therefore is the same as for the coherent state with the substitution $|\alpha|^4=\langle n\rangle^2$ by $N(N-1)/4$, i.e.
\begin{eqnarray}
\left \langle \mathbf  P^{(2)}(\vec \rho_1, \vec \rho_2  )  \right\rangle_\mathrm{\!coh2} \!\! &=& \! N(N\!-\!1) \cos^2 \!\left[\frac{k\, \ell}{2 z_0}  \rho_1 \right] \cos^2 \! \left[\frac{k\, \ell}{2 z_0}   \rho_2 \right]
\nonumber \\
&\,& \hspace*{-5 pt}\times
 \mathrm{sinc}^2 \!\left[  \frac{k\, a}{2 z_0}  \rho_1 \right]\mathrm{sinc}^2 \!\left[  \frac{k\, a}{2 z_0}  \rho_2 \right]
 \label{Eq:SM-2-photon-double-slit-cohN}
\end{eqnarray}
When the pattern is summed over $N$ with the proper weight coefficients, we see from the relation
\begin{eqnarray}
\sum_{N=0}^\infty N(N-1)\, |c^\alpha_N|^2
 \!=\!4\langle n\rangle^2
\end{eqnarray}
that the coherent result (\ref{Eq:SM-2-photon-coh-2-slit}) is obtained.

\subsection{Plots of the Second Order Patterns}
\label{SS:2nd-order-plots}

\begin{figure}[!h]
\centering
\includegraphics[width=1.0\columnwidth]{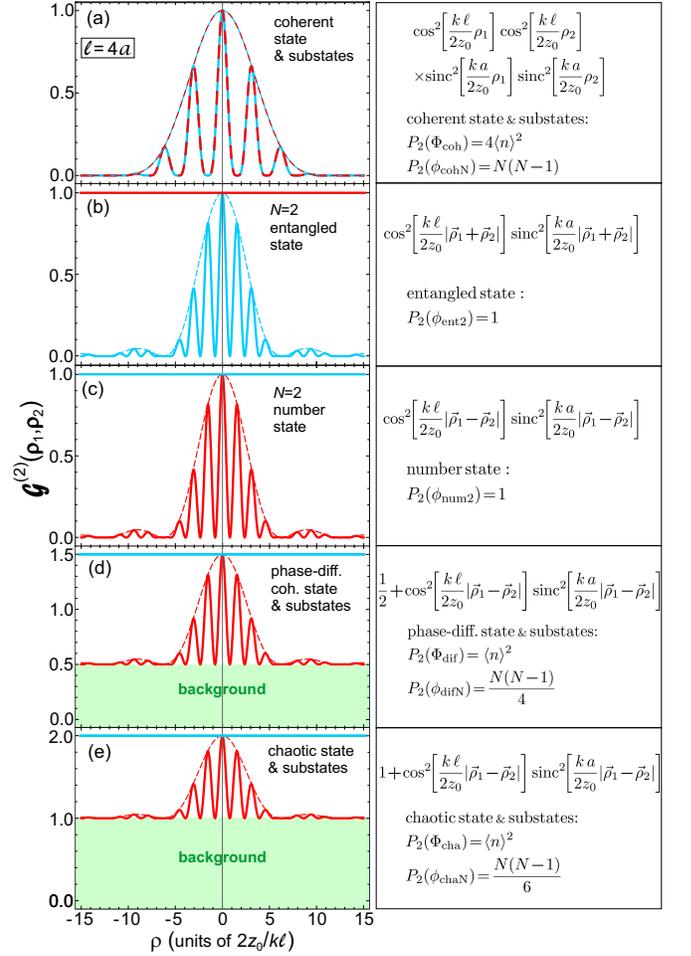}
\caption[]{Second order double-slit diffraction patterns for the indicated 2-mode quantum states and their substates for $\ell=4 a$. (a) Pattern $\boldsymbol {{\cal G}}^{(2)} (\vec \rho_1, \vec \rho_2  )$ for the coherent state and its binomial substates. (b) Pattern for the $N=2$ entangled  state, and (c) for the $N\!=\!2$ number  state. (d) Pattern for the phase-averaged coherent state and its substates, and (e) for the chaotic state and its substates.
}
 \label{Fig:2nd-ord-patterns}
\end{figure}

The calculated shapes $\boldsymbol {{\cal G}}^{(2)}  (\vec \rho_1, \vec \rho_2  )$  are plotted in Fig.\,\ref{Fig:2nd-ord-patterns} in the order of the cases in Fig.\,\ref{Fig:2-photon-sources}. Again the shapes and scaling factors  $P_2$ are given on the right for convenience. For the same quantum states, the degeneracy still present in the first order patterns in Fig.\,\ref{Fig:1st-ord-patterns}\,(b) is now lifted.  The diffraction patterns have become characteristic  signatures of the different quantum states of light, revealing the new paradigm.

The pattern in Fig.\,\ref{Fig:2nd-ord-patterns}\,(a) is characteristic of \emph{coherent states} emitted by the source. The shape is  independent of the total average number of photons, $2\langle n\rangle$, and the specific photon number, $N$, in the substates which only determine the scaling factor $P_2$.  The coherent states and their substates give the same patterns for the two detection schemes in Fig.\,\ref{Fig:geometry}\,(c).

For the specific source scheme on the right in Fig.\,\ref{Fig:2-photon-sources}\,(a), the source is described by a second order collective coherent state that contains an average number of $2\langle n\rangle =2$ photons. The collective state consists of a Poisson substate distribution extending up to about $N=10$ as shown in Fig.\,\ref{Fig:Probability-distributions}\,(a). The collective state may be imaged in two ways. The 2-photon coincidence detection scheme in Fig.\,\ref{Fig:geometry}\,(c) will pick out its representative $N=2$ substate defined by the arrival of photon pairs. On the other hand, a CCD detector will integrate over all arriving numbers of photons and thus record the entire collective state, as previously conjectured \cite{stohr:17,cloning-comment}. This is discussed in more detail in conjunction with the degree of second order coherence of the collective coherent state and its substates in the following section.

The patterns of the $N\!=\!2$ entangled state $|\phi_\mathrm{ent2}\rangle$  given by (\ref{Eq:SM-2-mode-ent-state}) and number state $|\phi_\mathrm{num2}\rangle$ expressed by (\ref{Eq:SM-2-mode-num-state}), are shown in  Figs.\,\ref{Fig:2nd-ord-patterns}\,(b) and (c). They are complete opposites, corresponding to the exchange of the red and blue detection schemes in Fig.\,\ref{Fig:geometry}\,(c). In particular, the well studied entangled state $|\phi_\mathrm{ent2}\rangle$ \cite{Ou:2007,Shih:2011,Ou:2017} incorporates the essence of quantum behavior since it is maximally entangled, and it plays a prominent role in quantum information science \cite{bouwmeester:2000,haroche:2001,horodecki:09,pan:2012}. The complementary behavior of the two states holds a central position in quantum optics since their diffraction patterns cannot be explained by the wave formalism.

The key role of the 2-photon entangled and number states is furthermore revealed by the fact that the 2-photon case of all five source implementations in Fig.\,\ref{Fig:2-photon-sources} may be represented as a linear combination of the two states, written in the general form
\begin{eqnarray}
&\,& \hspace*{-30pt} |\phi_{N=2} \rangle_{k,k'}
\nonumber \\
&\,&\hspace*{-30pt} = a_{11} \mathrm{e}^{\mathrm{i}\phi_{11}}  |1\rangle_k\, |1\rangle_{k'} + a_{02}\mathrm{e}^{\mathrm{i}\phi_{02}}  |0\rangle_k |2\rangle_{k'} +  a_{20} \mathrm{e}^{\mathrm{i}\phi_{20}} |2\rangle_k |0\rangle_{k'}
\hspace*{-15pt}
\nonumber \\
\label{Eq:SM-N-2-general}
\end{eqnarray}
where the coefficients $a_{ij}$ are real and for equal occupation of the 2-modes we have $a_{02} = a_{20}$.  The normalization condition furthermore links the coefficients by  $a_{11}^2 + a_{20}^2 + a_{02}^2=1$.

In particular, the $N\!=\!2$  coherent substate (\ref{Eq:SM-coh-N-bin-substates}), given by
\begin{eqnarray}
|\phi_\mathrm{coh2}\rangle_{k,k'}\!=   \frac{1 } {2} \! \left\{\sqrt{2}|1\rangle_k |1\rangle_{k'}\!  +\! \bigg[ |0\rangle_k| 2\rangle_{k'} \! + \! |2\rangle_k|0\rangle_{k'}\bigg] \!\right\}~~~
\label{Eq:SM-coh-2-bin-substates}
\end{eqnarray}
corresponds to an \emph{in-phase} addition of the $N=2$ state $|\phi_\mathrm{num2}\rangle$ and the state $|\phi_\mathrm{ent2}\rangle$,  identified by square brackets. Thus the coherent 2-photon pattern in Fig.\,\ref{Fig:2nd-ord-patterns}\,(a) arises from the interference of the number and entangled states, expressed by (\ref{Eq:SM-binomial-interference-term}).

The $N=2$ entangled  and number states are also responsible for the evolution of the patterns in Figs.\,\ref{Fig:2nd-ord-patterns}\,(c)--(e). The number state $|\phi_\mathrm{num2}\rangle=|1\rangle_k |1\rangle_{k'}$ creates the diffraction fine structure shown in (c) which is superposed on a background, shown in green, in the patterns of the collective phase-diffused  and chaotic states and their substates in (d) and (e). The size of the background in these patterns is determined by different relative contributions of the entangled and number states, which combine with random relative phases in the general 2-photon state (\ref{Eq:SM-N-2-general}). The key difference of the  $N=2$ phase-diffused coherent substate  (\ref{Eq:SM-dif-N-bin-substates}) and the chaotic substate (\ref{Eq:SM-cha-N-bin-substates}) is revealed by writing them respectively as,
\begin{eqnarray}
&\,& \hspace*{-40pt} |\phi_\mathrm{dif2}\rangle_{k,k'}\nonumber \\
&\,& \hspace*{-40pt} =\frac{1}{\sqrt{2}} \bigg\{ \mathrm{e}^{\mathrm{i}  \varphi }  |1\rangle_k |1\rangle_{k'} \! + \! \frac{1}{\sqrt{2}} \bigg[\mathrm{e}^{2\mathrm{i} \varphi } |0\rangle_k| 2\rangle_{k'} \! + \!  |2\rangle_k|0\rangle_{k'}\bigg ]\!\bigg\}
\label{Eq:SM-dif-2-bin-substates}
\end{eqnarray}
and
\begin{eqnarray}
&\,& \hspace*{-35pt} |\phi_\mathrm{cha2}\rangle_{k,k'}
\nonumber \\
&\,& \hspace*{-35pt} =\frac{1}{\sqrt{3}} \bigg \{\mathrm{e}^{  \mathrm{i}\varphi_{1}}  |1\rangle_k  |1\rangle_{k'}  + \bigg [ \mathrm{e}^{ \mathrm{i}\varphi_2} |0\rangle_k |2\rangle_{k'}  +     |2\rangle_k |0\rangle_{k'} \bigg] \!\bigg\}
\label{Eq:SM-cha-2-bin-substates}
\end{eqnarray}
In the last expression we have rewritten (\ref{Eq:SM-cha-N-bin-substates}) by eliminating an unimportant overall phase factor through the choice $\phi_{20}=0$, $\phi_{02}=\varphi_2$, and $\phi_{11}=\varphi_1$.

The different contributions of the entangled substate in (\ref{Eq:SM-dif-2-bin-substates}) and (\ref{Eq:SM-cha-2-bin-substates}), identified by rectangular brackets, relative to the number state is the reason for the change in the green background in the patterns in Fig.\,\ref{Fig:2nd-ord-patterns}\,(c)--(e). The background quantitatively scales with the square of the coefficients expressing the number and entangled state contributions to these states.

The entangled state reflects the simultaneous birth of two photons within a given slit. This situation is encountered in practice for incoherent or chaotic sources since the creation of only single photons per slit, reflected by a pure $|1\rangle_k |1\rangle_{k'}$ state (pattern in Fig.\,\ref{Fig:2nd-ord-patterns}\,(c)), requires special source preparation  \cite{santori:2002,patel:2010,flagg:2010,neuzner:2016}.

The dashed red envelope function of the chaotic state pattern in  Fig.\,\ref{Fig:2nd-ord-patterns}\,(e) is the 1D manifestation of the famous Hanbury Brown--Twiss (HBT) result \cite{Hanbury-Twiss:56a,Hanbury-Twiss:56,Hanbury-Twiss:57,Hanbury-Twiss:58,Hanbury-Twiss:58b}, where a circular or rectangular 2D source is replaced by a 1D slit of length $\ell$. The HBT effect was first derived quantum mechanically by Fano \cite{fano:61} using Feynman's concepts of probability amplitudes and played an important role in Glauber's development of quantum optics as recalled in his Nobel lecture \cite{glauber-nobel}.

In QED, the HBT arises naturally because of the structure of the chaotic quantum states. This is best revealed by the 2-photon chaotic substate (\ref{Eq:SM-cha-2-bin-substates}) which contains an entangled part of 50\%. It is therefore not surprising that many quantum optics experiments, first performed with entangled biphotons, can also be performed with chaotic light. A prominent example is ''ghost imaging''  \cite{strekalov:95,Shih:2011} which is also possible with chaotic sources \cite{bennink:02}, albeit with  reduced contrast \cite{valencia:2005}.

The HBT effect may also be explained classically as arising from ``intensity fluctuations'' \cite{purcell:56,wolf:2007} which of course are nothing but the fluctuations in the created number of photons per unit time and area. In the semi-classical picture, the constant background is typically attributed to 2-photons that ``accidentally'' arrive at the two detectors in coincidence. This arrival condition is assured for entangled biphotons created in parametric down conversion by their simultaneous birth at the same place. It may occur for a chaotic source when two photons born at different times and positions ``accidentally'' arrive at the same time because the difference in birth time is compensated by the difference in travel time (distance) \cite{stohr-AOP}. From a practical or detection point of view, the entangled and  ``accidental'' scenarios are indistinguishable. The formal statistical optics derivation utilizes the so-called Reed theorem \cite{Reed:62} or complex Gaussian moment theorem  \cite{Mandel-Wolf:65,Goodman-SO}, which veils the underlying fundamental quantum processes.

\subsection{Degree of Second Order Coherence}
\label{SS:2nd-order-coherence}

The difference of the second order patterns in Fig.\,\ref{Fig:2nd-ord-patterns} is also reflected by the degree of second order coherence (\ref{Eq:SM-Degree-of-coherence}) of the respective quantum states, which for the double slit geometry is expressed by,
\begin{eqnarray}
g^{(2)}(\vec \rho,-\vec \rho) =\frac{\left \langle \mathbf  P^{(2)}(\vec \rho, -\vec \rho  )  \right\rangle}
{ \left \langle \mathbf  P^{(1)}(\vec \rho, \vec \rho  )  \right\rangle^2  }
\label{Eq:Second-order-degree-of-coherence}
\end{eqnarray}
It is evaluated in Appendix C for the different cases in Fig.\,\ref{Fig:2nd-ord-patterns}.

For the \emph{collective} coherent state $|\Phi_\mathrm{coh}\rangle$, we have $g^{(2)}_\mathrm{coh}(\vec \rho,-\vec \rho)\!=\! 1$ which together with $g^{(1)}_\mathrm{coh}(\vec \rho,-\vec \rho)\!=\!1$ given by (\ref{Eq:g1-coh}) is the signature of a second order coherent state \cite{glauber:65,glauber-titulaer:65}. It is remarkable that the coherent \emph{substates}  $|\phi_\mathrm{cohN}\rangle$ are \emph{not} second order coherent since $g^{(2)}_\mathrm{cohN}(\vec \rho,-\vec \rho)=  1-\frac{1}{N}$, which approaches the coherent value of unity only in the limit of a large number of photons in the substates.

In particular, we obtain $g^{(2)}_\mathrm{coh2}(\vec \rho,-\vec \rho)=1/2$ for the 2-photon coherent substate  $|\phi_\mathrm{coh2}\rangle$ given by (\ref{Eq:SM-coh-2-bin-substates}). This state, describing a \emph{coherent biphoton} \cite{stohr:17,stohr-AOP}, is created when a single photon clones itself in a stimulated decay process. Its lack of second order coherence, which was previously not recognized \cite{cloning-comment} arises from the fact that in the presence of a single photon, an atom may also decay with equal probability via spontaneous photon emission. This is expressed by the well known factor $1+n$, where 1 is the relative probability that an excited electronic state decays \emph{spontaneously} in the absence of other photons, and $n$ is the relative probability that the decay is \emph{stimulated} by the presence of $n$ photons in the same mode.

The lack of second order coherence of the $N=2$ substate $|\phi_\mathrm{coh2}\rangle$, produced through single photon ($n=1$) stimulation, is  reflected by the so-called ``no cloning'' theorem \cite{Dieks:82, Milonni-Hardies:82, Wootters-Zurek:82, Mandel:83,lamas:2002}. Second order coherence and perfect cloning is only reached when many photons in the same mode cooperate to completely control excitation and de-excitation of an atom \cite{stohr-scherz:15}. This corresponds to the case shown on the right in Fig.\,\ref{Fig:2-photon-sources}\,(a) where an incident temporally coherent pulse with high degeneracy parameter drives the atoms in the film to a collective second order coherent state with equal populations in the ground and excited states. The coupled atom-photon system then becomes second order coherent and the ``no cloning'' theorem no longer applies.

The 2-photon entangled state $|\phi_\mathrm{ent2}\rangle$, given by (\ref{Eq:SM-2-mode-ent-state}), yields the surprising result $g^{(2)}_\mathrm{ent2}(\vec \rho,-\vec \rho)= 1$. A state is second order coherent, however, only if both $g^{(2)}$ and $g^{(1)}$ are unity. This is not the case because  $g^{(1)}$ is not unity according to (\ref{g1-incoherent-states}). Instead, the state $|\phi_\mathrm{ent2}\rangle$  causes the constant background in Figs.\,\ref{Fig:2nd-ord-patterns}\,(d) and (e) as discussed in the previous section. The unit value of the background is therefore deceiving since it does not reflect second order coherence, which in the semiclassical explanation has led to its description as ``accidental coincidences''.

For the 2-photon number state $|\phi_\mathrm{num2}\rangle$, the collective phase-diffused coherent state $|\Phi_\mathrm{dif}\rangle$ and the collective chaotic state $|\Phi_\mathrm{dif}\rangle$, we find $g^{(2)} (\vec \rho,-\vec \rho)=\boldsymbol {{\cal G}}^{(2)}  (\vec \rho,-\vec \rho)$. The red patterns in Fig.\,\ref{Fig:2nd-ord-patterns}\,(c)--(e) therefore represent the degree of second order coherence of these states. For the substates  $|\phi_\mathrm{difN}\rangle$ and $|\phi_\mathrm{chaN}\rangle$ the expressions for $g^{(2)}$, given by (\ref{Eq:g2-difN-state}) and (\ref{Eq:g2-chaN-state}) in Appendix C, are similar but contain additional prefactors.

The values of the degree of second order coherence $g^{(2)} (\vec \rho_1, \vec \rho_2  )$  of the different quantum states complements the  information revealed by the shapes of their diffraction patterns $\boldsymbol {{\cal G}}^{(2)} (\vec \rho_1, \vec \rho_2  )$ in Fig.\,\ref{Fig:2nd-ord-patterns}. While the patterns of the collective states $\left \langle \Phi_\mathrm{i} |\mathbf  P^{(2)}(\vec \rho_1, \vec \rho_2  )|\Phi_\mathrm{i} \right \rangle $ and their substates $\left \langle \phi_\mathrm{iN} |\mathbf  P^{(2)}(\vec \rho_1, \vec \rho_2  )|\phi_\mathrm{iN} \right \rangle $ can in principle be distinguished through their scaling factors $P_2$, their difference is directly revealed by the normalized degree of second order coherence $g^{(2)} (\vec \rho_1, \vec \rho_2  )$. Examples are the different values of $g^{(2)}_\mathrm{coh} = 1$ for the collective coherent state and $g^{(2)}_\mathrm{coh2} = 1/2$ for its 2-photon substate which exhibit the same diffraction shapes.

\subsection{The Evolution from First to Second Order}

The behavior of independent photons in first order QED may also be accounted for by the wave theory, augmented by certain \emph{ad hoc} recipes like the Huygens-Fresnel principle which make it work. In second order QED, phenomena arise that simply cannot be explained by the wave theory of light, clearly revealing its incompleteness. The hallmark of second order QED is the existence of correlations between photons, the most heralded phenomenon being  photon \emph{entanglement} over large distances \cite{horodecki:09,pan:2012}.  More precisely, the second order case covers phenomena associated with a number of $N\!\geq\!2$ photons that arise from the correlations between their probability amplitudes. These correlations are absent in first order QED or conventional quantum mechanics, as expressed by Dirac's famous statement.

Comparison of the first order patterns in  Fig.\,\ref{Fig:1st-ord-patterns} with the second order ones in Fig.\,\ref{Fig:2nd-ord-patterns} directly reveals how the remaining degeneracy in the first order patterns is lifted in second order. In particular, the evolution  leads to distinct patterns for the fundamental 2-mode entangled and number states, and their central role becomes apparent.

In all cases, the shapes of the second order diffraction profiles (apart from any constant background) are seen to be the \emph{square} of the corresponding first order patterns. For the coherent states the \emph{effective width} of the first order pattern for $\ell \geq 2a$ is given by \cite{stohr-AOP}
\begin{eqnarray}
\frac{k a }{ \pi z_0} \int_{-\infty}^\infty  \cos^2\! \left[ \frac{k\, \ell}{2 z_0} \rho \right] \mathrm{sinc}^2 \left[  \frac{k\, a}{ 2 z_0} \rho\right]\,\mathrm{d} \rho=1
\label{Eq:SM-1st-order-integration}
\end{eqnarray}
while that of the second order coherent pattern is given by
\begin{eqnarray}
\frac{ k a }{ \pi z_0}\int_{-\infty}^\infty  \cos^4\! \left[ \frac{k\, \ell}{2 z_0} \rho \right] \mathrm{sinc}^4 \left[  \frac{k\, a}{ 2 z_0} \rho\right]\,\mathrm{d} \rho  = \frac{1}{2}
\end{eqnarray}
Photon conservation then requires that the reduction in effective width of the second order pattern by a factor of 2 is compensated by a  factor of 2 larger peak value. This illustrates that with increasing order of coherence the pattern is increasingly centered around the forward direction. When extended to higher order this leads to the remarkable result that, in principle, an $n^{th}$ order coherent state no longer diffracts and the collective photon state propagates on particle-like trajectories \cite{stohr-AOP}.

\section{Summary and Conclusions}

The key message of our paper is that diffraction patterns can simply be viewed as encoded signatures of different quantum states of light. This is revealed by a quantum formulation of diffraction that goes beyond the first order description in conventional quantum mechanics. The theoretical link between quantum states and their characteristic diffraction images is revealed by modern versions of Young's double slit diffraction experiment, summarized in Fig.\,\ref{Fig:2-photon-sources}. Ironically, the very experiment that 200 years ago led to the notion that light is a wave, can therefore be used to disprove this hypothesis.

We note that the true photon nature of what we call electromagnetic (EM) radiation is not restricted to the short wavelength range extending from the optical to the x-ray regime. It is a universal feature of EM radiation despite the power of Maxwell's classical theory of electromagnetic waves. Owing to the lower energy of photons below the visible range, it just becomes increasingly difficult to detect them since typical detectors at room temperature have a thermal background noise corresponding to about 25\,meV. Today, even micowave photons may be detected by use of photon counters based on Josephson junctions \cite{chen:11}.

In principle, the wave theory can be abandoned altogether today. In practice, it has served us well and may continue to be used with the understanding of its limitations. The broad and difficult concept of ``partial coherence'' in statistical optics can be better defined in quantum optics, which is anchored in the fundamental theory of light and matter, QED. The definition of quantum coherence is more specific since it is directly linked to different quantum states of light whose interference and diffraction properties are quantified through an  order-dependent degree of coherence.  The quantum theory furthermore differentiates between the behavior of collective quantum states defined through statistical distributions of photons and states containing specific numbers of photons.

In \emph{first order} QED, the fundamental photon nature of light just happens to be describable by the \emph{ad hoc} wave theory, based on spherical light waves, the magical Huygens-Fresnel principle, and the assumption that the absolute value squared of the wave field gives the diffracted intensity. The limitations and non-fundamental nature of the wave theory become apparent only in \emph{second order} QED, where the concept of spherical waves needs to be replaced by an average over photon emission directions, the Huygens-Fresnel principle becomes the quantum interference of photon probability amplitudes for different source-detector paths, and the ``intensity'' concept is replaced by the probability of photon detection.

\subsection{Ackowledgements}

I would like to thank Jianbin Liu of Xi'an Jiaotong University for many elucidating discussions and comments.

\appendix

\section{Derivation of First Order Diffraction Patterns}

\subsection{Phase-Diffused Coherent State}

For the \emph{phase-diffused coherent state} $|\Phi_\mathrm{dif}\rangle$ the non-vanishing matrix elements  are given by
\begin{eqnarray}
&\,&\langle \Phi_\mathrm{dif} |\mathbf a^\dag_{\vec{k} } \, \mathbf   a_{ \vec{k} } |\Phi_\mathrm{dif}\rangle
=  \langle \Phi_\mathrm{dif} |\mathbf a^\dag_{\vec{k}' } \, \mathbf   a_{ \vec{k}' } |\Phi_\mathrm{dif}\rangle
\nonumber \\
&\,& = \sum_{N=0}^\infty \frac{ |\alpha|^{2N}}{\mathrm{e}^{ 2 |\alpha|^2}} \,  \sum_{m=0}^N  \,   \frac{m}{m!(N-m)!  }=|\alpha|^{2}
\label{Eq:SM-matrix-dif1}
\end{eqnarray}
and
\begin{eqnarray}
\langle \Phi_\mathrm{dif} |\mathbf a^\dag_{\vec{k} } \, \mathbf   a_{ \vec{k}' } |\Phi_\mathrm{dif}\rangle
&=&|\alpha|^{2} \,  \mathrm{e}^{ \mathrm{i}\varphi}
\nonumber \\
 \langle \Phi_\mathrm{dif} |\mathbf a^\dag_{\vec{k}' } \, \mathbf   a_{ \vec{k}  } |\Phi_\mathrm{dif}\rangle
&=&|\alpha|^{2} \,  \mathrm{e}^{ -\mathrm{i}\varphi}
\label{Eq:SM-matrix-dif2}
\end{eqnarray}

The $\mathbf Y$-term vanishes for a phase average over $\varphi$. For our assumed geometry in Fig.\,1\,(b)  we obtain in the $(\vec r,\vec \rho)$ coordinates and by integration over the slit width $a$ and with $\langle n  \rangle=|\alpha|^2$,
\begin{eqnarray}
 \left \langle   \mathbf  P^{(1)}(\vec \rho_1,   \vec \rho_2  )  \right\rangle_\mathrm{\!dif} \!&=&\!  \langle n  \rangle \cos\! \left[ \frac{k\, \ell}{2 z_0} |\vec \rho_1-\vec \rho_2| \right]
 \nonumber \\
 &\,& \times \mathrm{sinc} \left[  \frac{k\, a}{ 2 z_0} |\vec \rho_1-\vec \rho_2|\right]
\label{Eq:SM-1-photon-avg-pattern}
\end{eqnarray}

\subsection{$N$-Photon Substate of Phase-Diffused Coherent State}

For the substate $|\phi_\mathrm{difN} \rangle$ of the phase-diffused coherent state, the matrix elements are given by (\ref{Eq:SM-matrix-dif1}) and (\ref{Eq:SM-matrix-dif2}) with $|\alpha|^{2}$ replaced by $N/2$. The pattern is obtained as,
\begin{eqnarray}
\left \langle   \mathbf  P^{(1)}(\vec \rho_1,   \vec \rho_2  )  \right\rangle_{\mathrm{difN} }  \!&=&\! \frac{N}{2}   \cos\! \left[ \frac{k\, \ell}{2 z_0} |\vec \rho_1-\vec \rho_2| \right]
\nonumber \\
 &\,& \times \mathrm{sinc} \left[  \frac{k\, a}{ 2 z_0} |\vec \rho_1-\vec \rho_2|\right]
\label{Eq:SM-1P-chiN-pattern}
\end{eqnarray}
When summed over $N$ with the weight factor  $ |c^\alpha_N|^2$ it is seen from the relation
\begin{eqnarray}
\sum_{N=0}^\infty |c^\alpha_N|^2 \frac{N }{2} =  |\alpha|^2=\langle n\rangle
\end{eqnarray}
that (\ref{Eq:SM-1P-chiN-pattern}) becomes the pattern of the phase-diffused state given by (\ref{Eq:SM-1-photon-avg-pattern}), as required.

\subsection{Chaotic State}

The non-vanishing matrix elements of the  \emph{ 2-mode chaotic state} $|\Phi_\mathrm{cha}\rangle$ are given by
\begin{eqnarray}
\langle \Phi_\mathrm{cha}|\mathbf  a_k^{\dagger}  \mathbf  a_k   |\Phi_\mathrm{cha} \rangle \! &=&  \! \langle \Phi_\mathrm{cha}|\mathbf  a_{k'}^{\dagger}  \mathbf  a_{k'}   |\Phi_\mathrm{cha} \rangle
\nonumber \\
\! &=&  \! \sum_{N=0}^\infty\frac{N  \,\langle n  \rangle^N}{\left( 1\! + \!\langle n
\rangle \right)^{N + 1}}  \! =  \! \langle n  \rangle
\label{Eq:SM-matrix-cha}
 \end{eqnarray}
since all other matrix element contain phase factors that average to zero. The detection probability (1) is therefore the same as for the phase-diffused coherent state and we have
\begin{eqnarray}
\left \langle   \mathbf  P^{(1)}(\vec \rho_1,   \vec \rho_2  )  \right\rangle_\mathrm{\!cha} \!&=&\! \langle n  \rangle \cos\! \left[ \frac{k\, \ell}{2 z_0} |\vec \rho_1-\vec \rho_2| \right]
 \nonumber \\
 &\,& \times
\mathrm{sinc} \left[  \frac{k\, a}{ 2 z_0} |\vec \rho_1-\vec \rho_2|\right]
\label{Eq:SM-1-photon-cha-pattern}
\end{eqnarray}

\subsection{$N$-Photon Substate  of Chaotic State}

For the chaotic substate $|\phi_\mathrm{chaN} \rangle$ the matrix elements are given by (\ref{Eq:SM-matrix-cha})  with $\langle n\rangle $ replaced by $N/2$. The pattern is obtained as,
\begin{eqnarray}
\left \langle   \mathbf  P^{(1)}(\vec \rho_1,   \vec \rho_2  )  \right\rangle_{\mathrm{chaN} }\! &=&\!  \frac{N}{2} \cos\! \left[ \frac{k\, \ell}{2 z_0} |\vec \rho_1-\vec \rho_2| \right]  \nonumber \\
 &\,& \times
 \mathrm{sinc} \left[  \frac{k\, a}{ 2 z_0} |\vec \rho_1-\vec \rho_2|\right]
\label{Eq:SM-1P-psiN-pattern}
\end{eqnarray}
If we sum over $N$ with the weight factors $|c^\beta_N|^2$ we obtain by use of
\begin{eqnarray}
\sum_{N=0}^\infty  |c^\beta_N|^2 \frac{N}{2} =   \langle n \rangle
\end{eqnarray}
the chaotic result (\ref{Eq:SM-1-photon-cha-pattern}).

\subsection{$N$-Photon Entangled (NOON) State}

For the \emph{2-mode $N$-photon entangled state} $|\phi_\mathrm{entN}\rangle$, the  matrix elements are
\begin{eqnarray}
 \langle \phi_\mathrm{entN} |\mathbf a^\dag_{\vec{k} } \, \mathbf   a_{ \vec{k}' } |\phi_\mathrm{entN}\rangle \! &=& \! \langle \phi_\mathrm{entN} |\mathbf a^\dag_{\vec{k}' } \, \mathbf   a_{ \vec{k}  } |\phi_\mathrm{entN}\rangle  =0
\nonumber\\
 \langle \phi_\mathrm{entN} |\mathbf a^\dag_{\vec{k} } \, \mathbf   a_{ \vec{k}  } |\phi_\mathrm{entN}\rangle \! &=& \! \langle \phi_\mathrm{entN} |\mathbf a^\dag_{\vec{k}' } \, \mathbf   a_{ \vec{k}'  } |\phi_\mathrm{entN}\rangle  =\frac{N}{2}
 \hspace*{-15pt}
 \nonumber\\
 \end{eqnarray}
We obtain
\begin{eqnarray}
\left \langle   \mathbf  P^{(1)}(\vec \rho_1,   \vec \rho_2  )  \right\rangle_\mathrm{\!entN} \! &=&\!  \frac{N}{2}    \cos\! \left[ \frac{k\, \ell}{2 z_0} |\vec \rho_1-\vec \rho_2| \right]
\nonumber \\
 &\,& \times  \mathrm{sinc} \left[  \frac{k\, a}{ 2 z_0} |\vec \rho_1-\vec \rho_2|\right]
\label{Eq:SM-1P-ent-pattern}
\end{eqnarray}

\subsection{$N$-Photon Number State}
\label{SS:number-state}

For the \emph{2-mode $N$-photon number state} $|\phi_\mathrm{numN}\rangle$  with $N\geq 2$ the matrix elements are evaluated as
\begin{eqnarray}
 \langle \phi_\mathrm{numN} |\mathbf a^\dag_{\vec{k} } \, \mathbf   a_{ \vec{k}' } |\phi_\mathrm{numN}\rangle \! \! &=&\!\! \langle \phi_\mathrm{numN} |\mathbf a^\dag_{\vec{k}' } \, \mathbf   a_{ \vec{k}  } |\phi_\mathrm{numN}\rangle  =0
\nonumber\\
 \langle \phi_\mathrm{numN} |\mathbf a^\dag_{\vec{k} } \, \mathbf   a_{ \vec{k}  } |\phi_\mathrm{numN}\rangle \! \! &=&\! \! \langle \phi_\mathrm{numN} |\mathbf a^\dag_{\vec{k}' } \, \mathbf   a_{ \vec{k}'  } |\phi_\mathrm{numN}\rangle  =\frac{N}{2}
  \hspace*{-15pt}
 \nonumber\\
\end{eqnarray}
which is the same as for the entangled state (\ref{Eq:SM-1P-ent-pattern}).

\section{Derivation of Second Order Diffraction Patterns}

\subsection{Phase-Diffused Coherent State}

For the \emph{phase-diffused coherent state} $|\Phi_\mathrm{dif}\rangle$ the  matrix elements for term $\langle \mathbf A \rangle$ are obtained as,
\begin{eqnarray}
&\,& \langle \Phi_\mathrm{dif} |  \mathbf a_{\vec{k}}^{\dagger} \,\mathbf  a_{ \vec{k}'}^{\dagger} \, \mathbf a_{\vec{k} }  \,   \mathbf  a_{ \vec{k}'} |\Phi_\mathrm{dif}\rangle  = \langle \Phi_\mathrm{dif} |  \mathbf a_{\vec{k}'}^{\dagger} \,\mathbf  a_{ \vec{k}}^{\dagger} \, \mathbf a_{\vec{k}' }  \,   \mathbf  a_{ \vec{k} } |\Phi_\mathrm{dif}\rangle
 \nonumber \\
 &\,& = \sum_{N=0}^\infty \frac{|\alpha|^{2N}  }{\mathrm{e}^{2|\alpha|^2}} \sum_{m=0}^N \frac{m(m-1)}{m!(N-m)!} = |\alpha|^4
\label{Eq:term-A-dif}
\end{eqnarray}

For the first two terms in $\langle \mathbf B \rangle$ we have,
\begin{eqnarray}
\langle \Phi_\mathrm{dif} |  \mathbf a_{\vec{k}}^{\dagger} \,\mathbf  a_{ \vec{k} }^{\dagger} \, \mathbf a_{\vec{k} }  \,   \mathbf  a_{ \vec{k}} |\Phi_\mathrm{dif}\rangle \!&=&\!\langle \Phi_\mathrm{dif} |  \mathbf a_{\vec{k}'}^{\dagger} \,\mathbf  a_{ \vec{k} '}^{\dagger} \, \mathbf a_{\vec{k}' }  \,   \mathbf  a_{ \vec{k}'} |\Phi_\mathrm{dif}\rangle
\nonumber \\
&=&\! |\alpha|^4
\label{Eq:term-B-dif-1}
\end{eqnarray}
while for the last two terms in  $\langle \mathbf B \rangle$  we obtain,
\begin{eqnarray}
\langle \Phi_\mathrm{dif} |  \mathbf a_{\vec{k}}^{\dagger} \,\mathbf  a_{ \vec{k} }^{\dagger} \, \mathbf a_{\vec{k}' }  \,   \mathbf  a_{ \vec{k}'} |\Phi_\mathrm{dif}\rangle \!&=&\! \mathrm{e}^{ 2 \mathrm{i}  \varphi}  \, |\alpha|^4
\nonumber \\
\langle \Phi_\mathrm{dif} |  \mathbf a_{\vec{k}'}^{\dagger} \,\mathbf  a_{ \vec{k}' }^{\dagger} \, \mathbf a_{\vec{k} }  \,   \mathbf  a_{ \vec{k} } |\Phi_\mathrm{dif}\rangle \!&=&\! \mathrm{e}^{- 2 \mathrm{i}  \varphi}  \, |\alpha|^4
\label{Eq:term-B-dif-2}
\end{eqnarray}
The terms $\langle \mathbf C \rangle$ and $\langle \mathbf B \rangle$ are evaluated as,
\begin{eqnarray}
 \langle \Phi_\mathrm{dif} |  \mathbf a_{\vec{k}}^{\dagger} \,\mathbf  a_{ \vec{k} }^{\dagger} \, \mathbf a_{\vec{k}  }  \,   \mathbf  a_{ \vec{k}'} |\Phi_\mathrm{dif}\rangle &=& \langle \Phi_\mathrm{dif} |  \mathbf a_{\vec{k}}^{\dagger} \,\mathbf  a_{ \vec{k}' }^{\dagger} \, \mathbf a_{\vec{k}'}  \,   \mathbf  a_{ \vec{k}'} |\Phi_\mathrm{dif}\rangle
 \nonumber \\
 &=& \mathrm{e}^{   \mathrm{i}  \varphi}   \, |\alpha|^4
\nonumber\\
\langle \Phi_\mathrm{dif} |  \mathbf a_{\vec{k}'}^{\dagger} \,\mathbf  a_{ \vec{k} }^{\dagger} \, \mathbf a_{\vec{k} }  \,   \mathbf  a_{ \vec{k}} |\Phi_\mathrm{dif}\rangle
&=&\langle \Phi_\mathrm{dif} |  \mathbf a_{\vec{k}'}^{\dagger} \,\mathbf  a_{ \vec{k}' }^{\dagger} \, \mathbf a_{\vec{k} ' }  \,   \mathbf  a_{ \vec{k} } |\Phi_\mathrm{dif}\rangle
 \nonumber \\
&=& \mathrm{e}^{  - \mathrm{i}  \varphi}   \, |\alpha|^4
\end{eqnarray}
All terms containing the phase $\varphi$ vanish upon phase averaging and we obtain with $|\alpha|^2=\langle n \rangle$,
\begin{eqnarray}
\left \langle \mathbf  P^{(2)}(\vec x_1, \vec x_2  )  \right\rangle_\mathrm{\!dif}\!
= \! \langle n \rangle^2 \left\{ \frac{1}{2} \! + \!
\cos^2 \!\left [ \frac{1}{2}  (\vec{k} \! - \! \vec{k}') \cdot (\vec{x}_1 \! - \! \vec{x}_2)\right ] \! \right\}
\hspace*{-15pt}
\nonumber \\
\label{Eq:phase-dif-coh-state-pattern}
\end{eqnarray}
When expressed in the coordinates $(\vec r,\vec \rho)$ and integrated over the slit width $a$  we obtain,
\begin{eqnarray}
  \hspace*{-25pt} \left \langle \mathbf  P^{(2)}(\vec \rho_1, \vec \rho_2  )  \right\rangle_\mathrm{\!dif}
 \! &=& \! \langle n \rangle^2\bigg\{ \frac{1}{2} \!+\! \cos^2 \! \left[\frac{k\, \ell}{ 2 z_0} |\vec \rho_1 \! -\! \vec \rho_2|  \right ]
\nonumber \\
&\,& \times
 \mathrm{sinc}^2 \!\left[  \frac{k\, a}{2 z_0} |\vec \rho_1 \! - \! \vec \rho_2|\right] \!\bigg\}
 \label{Eq:phase-dif-coh-state-pattern-2}
\end{eqnarray}
The pattern is  constant, $\left \langle \mathbf  P^{(2)}(\vec \rho_1, \vec \rho_2  )  \right\rangle_\mathrm{dif}= 3\langle n \rangle^2/2$ for the detection geometry $\vec \rho_1 \! =\! \vec \rho_2$ and for $\vec \rho_1 \! =\! -\vec \rho_2$ the diffraction fine structure sits on a constant background $\langle n \rangle^2/2$.

\subsection{$N$-Photon Substate of Phase-Diffused Coherent State}

For the substate $|\phi_\mathrm{difN} \rangle$ the matrix elements are the same as those of the collective parent state with $|\alpha|^4$ replaced by $N(N-1)/4$ so that the pattern is,
\begin{eqnarray}
 \left \langle \mathbf  P^{(2)}(\vec \rho_1, \vec \rho_2  )  \right\rangle_\mathrm{\mathrm{\!difN}}
\! &=& \! \frac{N(N-1)}{4 }  \bigg\{ \frac{1}{2} \!+\! \cos^2 \! \left[\frac{k\, \ell}{ 2 z_0} |\vec \rho_1 \! -\! \vec \rho_2|  \right ]
\nonumber \\
&\,& \times \mathrm{sinc}^2 \!\left[  \frac{k\, a}{2 z_0} |\vec \rho_1 \! - \! \vec \rho_2|\right] \!\bigg\}
 \label{Eq:SM-phase-avg-coh-state-pattern-2}
\end{eqnarray}
When summed over $N$ with the weight factors  $|c^\alpha_N|^2$ it is seen from the relation
\begin{eqnarray}
\sum_{N=0}^\infty\frac{N(N-1)}{4 } \, |c^\alpha_N|^2 =  |\alpha|^4=\langle n \rangle^2
\end{eqnarray}
that (\ref{Eq:SM-phase-avg-coh-state-pattern-2}) becomes the pattern of the phase-diffused state given by (\ref{Eq:phase-dif-coh-state-pattern-2}), as required.

\subsection{Chaotic State}

For the  \emph{ 2-mode chaotic state} $|\Phi_\mathrm{cha}\rangle$ the terms in  $\langle  \mathbf  A\rangle$  are evaluated as,
\begin{eqnarray}
&\,& \langle \Phi_\mathrm{cha} | \mathbf a_{\vec{k}}^{\dagger} \,\mathbf  a_{ \vec{k}'}^{\dagger} \, \mathbf a_{\vec{k} }  \,   \mathbf  a_{ \vec{k}'}  |\Phi_\mathrm{cha} \rangle =\langle \Phi_\mathrm{cha} | \mathbf a_{\vec{k}'}^{\dagger} \,\mathbf  a_{ \vec{k}}^{\dagger} \, \mathbf a_{\vec{k}' }  \,   \mathbf  a_{ \vec{k}}  |\Phi_\mathrm{cha} \rangle
\nonumber \\
 &\,& = \bigg [\sum_{N=0}^\infty  N  \,\frac{\langle n  \rangle^{N}}{\left( 1\! + \!\langle n
\rangle \right)^{N + 1}}\bigg] ^2
\nonumber \\
 &\,& =  \sum_{N=0}^\infty   \sum_{m=0}^N  m(N-m)  \,\frac{\langle n  \rangle^{N}}{\left( 1\! + \!\langle n
\rangle \right)^{N + 2}}
 =   \langle n  \rangle^2
 \label{Eq:SM-cha-exp-value-n2}
 \end{eqnarray}
Similarly, the contributions of the first two terms of $\langle  \mathbf B\rangle$  are obtained as,
\begin{eqnarray}
 &\,& \hspace*{-30pt} \langle \Phi_\mathrm{cha} | \mathbf a_{\vec{k}}^{\dagger} \,\mathbf  a_{ \vec{k} }^{\dagger} \, \mathbf a_{\vec{k} }  \,   \mathbf  a_{ \vec{k} }|\Phi_\mathrm{cha} \rangle =\langle \Phi_\mathrm{cha} | \mathbf a_{\vec{k}'}^{\dagger} \,\mathbf  a_{ \vec{k}' }^{\dagger} \, \mathbf a_{\vec{k}' }  \,   \mathbf  a_{ \vec{k}' }|\Phi_\mathrm{cha} \rangle
 \nonumber \\ &\,& = \sum_{N=0}^\infty  \sum_{m=0}^N  m(m-1) \frac{\langle n  \rangle^{N}}{\left( 1\! + \!\langle n  \rangle \right)^{N + 2}}\!
=\!  2 \langle n  \rangle^2
\label{Eq:SM-term-B-chaotic-case}
\end{eqnarray}
while the contributions from the other two terms in $\langle  \mathbf B\rangle$  average to zero, i.e.
\begin{eqnarray}
 \langle \Phi_\mathrm{cha} | \mathbf a_{\vec{k} }^{\dagger}      \mathbf a_{\vec{k} }^{\dagger}   \mathbf  a_{ \vec{k}'}   \mathbf  a_{ \vec{k}'}|\Phi_\mathrm{cha} \rangle \! = \!  \langle \Phi_\mathrm{cha} | \mathbf  a_{ \vec{k}'}^{\dagger}   \mathbf  a_{ \vec{k}'}^{\dagger} \mathbf a_{\vec{k} }    \mathbf a_{\vec{k} } |\Phi_\mathrm{cha} \rangle \! = \!0
  \hspace*{-15pt}
 \nonumber\\
 \label{Eq:SM-term-B-zero-average}
\end{eqnarray}
The contribution by the term $ \langle  \mathbf B \rangle$ is therefore only a constant resulting in a background. The matrix elements associated with  terms $\langle  \mathbf  C \rangle $ and $\langle \mathbf D \rangle$  are zero owing to the fact that they contain unpaired raising and lowering operators, i.e. $\langle  \mathbf  C+  \mathbf D \rangle=0$. Evaluation of detection probability and phase averaging yields,
\begin{eqnarray}
\left \langle \mathbf  P^{(2)}(\vec x_1, \vec x_2  )  \right\rangle_\mathrm{\!cha} \!=\! \langle n  \rangle^2\bigg\{ 1 \!+ \! \cos^2 \! \bigg [\frac{1}{2}(\vec{k} \!- \!\vec{k}')\cdot(\vec{x}_1 \!- \!\vec{x}_2)\bigg] \! \bigg\}
\hspace*{-15pt}
\nonumber \\
\end{eqnarray}
In the coordinates $(\vec r,\vec \rho)$ and integration over the slit width $a$ we obtain,
\begin{eqnarray}
\hspace*{-15pt} \left \langle \mathbf  P^{(2)}(\vec \rho_1, \vec \rho_2  )  \right\rangle_\mathrm{\!cha}
\!\! &=&\! \langle n  \rangle^2 \bigg\{ 1 \!+\! \cos^2 \! \left[\frac{k\, \ell}{ 2 z_0} |\vec \rho_1 \! -\! \vec \rho_2|  \right ]
 \nonumber \\
 &\,& \times \mathrm{sinc}^2 \!\left[  \frac{k\, a}{2 z_0} |\vec \rho_1 \! - \! \vec \rho_2|\right ] \! \bigg\}
\label{Eq:SM-Prob-2-mode-cha-state}
\end{eqnarray}

\subsection{$N$-Photon Substate of Chaotic State}

For the chaotic substate $|\phi_\mathrm{chaN} \rangle$   the  matrix elements are those of the collective parent state with  $\langle n  \rangle^2$ replaced by $N(N\!-\!1)/6$ and we obtain,
\begin{eqnarray}
\hspace*{-15pt} \left \langle \mathbf  P^{(2)}(\vec \rho_1, \vec \rho_2  )  \right\rangle_{\mathrm{\!chaN}}
\! \!&=&\! \frac{N(N\!-\!1)}{6}  \bigg\{ 1 \!+\! \cos^2 \! \left[\frac{k\, \ell}{ 2 z_0} |\vec \rho_1 \! -\! \vec \rho_2|  \right ]
 \nonumber \\
 &\,& \times \mathrm{sinc}^2 \!\left[  \frac{k\, a}{2 z_0} |\vec \rho_1 \! - \! \vec \rho_2|\right ]\bigg\}
\label{Eq:SM-Prob-chaN-state}
\end{eqnarray}
By summing (\ref{Eq:SM-Prob-chaN-state}) over $N$ with the weight factors $|c^\beta_N|^2$ we obtain by use of the relation
\begin{eqnarray}
\sum_{N=0}^\infty |c^\beta_N|^2\frac{N(N\!-\!1)}{6} = \langle n  \rangle^2
 \end{eqnarray}
the chaotic result (\ref{Eq:SM-Prob-2-mode-cha-state}), as required.

\subsection{$N$-Photon Entangled (NOON) State}

For the \emph{2-mode $N$-photon entangled state} $|\phi_\mathrm{entN}\rangle$  the matrix elements for $\mathbf A$ vanish,
\begin{eqnarray}
 \langle \phi_\mathrm{entN} |\mathbf a^\dag_{\vec{k} } \, \mathbf   a^\dag_{ \vec{k}'} \mathbf a_{\vec{k}' } \, \mathbf   a_{ \vec{k}} |\phi_\mathrm{entN}\rangle =0
\end{eqnarray}
The matrix elements for the first two terms in $\mathbf B$ are
\begin{eqnarray}
 \langle \phi_\mathrm{entN} |\mathbf a^\dag_{\vec{k} } \, \mathbf   a^\dag_{ \vec{k} } \mathbf a_{\vec{k}  } \, \mathbf   a_{ \vec{k}} |\phi_\mathrm{entN}\rangle \!&=&\! \langle \phi_\mathrm{entN} |\mathbf a^\dag_{\vec{k}' } \, \mathbf   a^\dag_{ \vec{k}' } \mathbf a_{\vec{k}  ' } \, \mathbf   a_{ \vec{k}'} |\phi_\mathrm{entN}\rangle
 \nonumber \\
 \!&=&\! \frac{N(N-1)}{2}
\end{eqnarray}
while those for the second two terms in $\mathbf B$ are non-zero only for $N=2$,
\begin{eqnarray}
 \langle\phi_\mathrm{entN} |\mathbf a^\dag_{\vec{k} } \, \mathbf   a^\dag_{ \vec{k} } \mathbf a_{\vec{k}'  } \, \mathbf   a_{ \vec{k}'} |\phi_\mathrm{entN}\rangle \!&=&\! \langle\phi_\mathrm{entN} |\mathbf a^\dag_{\vec{k}' } \, \mathbf   a^\dag_{ \vec{k} '} \mathbf a_{\vec{k}  } \, \mathbf   a_{ \vec{k} } |\phi_\mathrm{entN}\rangle
 \nonumber \\  \!&=&\! \delta(N,2)
\end{eqnarray}
where $\delta(N,2)=1$ for $N=2$ and zero otherwise.

The diffraction pattern is a constant for $N > 2$, given by $ \left \langle \mathbf  P^{(2)}(\vec x_1, \vec x_2  )  \right\rangle_\mathrm{entN}=  \frac{N(N-1)}{4}  $. For the specific $N=2$ case we obtain,
\begin{eqnarray}
 \hspace*{-20pt} \left \langle \mathbf  P^{(2)}(\vec x_1, \vec x_2  )  \right\rangle_\mathrm{\!ent2} \!\!&=& \!\! \frac{1}{4}\left \langle  \phi_\mathrm{ent2} |\mathbf B|  \phi_\mathrm{ent2} \right \rangle \nonumber \\
\!\!&=& \!\!\cos^2\left [ \frac{1}{2}(\vec{k} \! - \! \vec{k}') \cdot (\vec {x}_1 \! + \! \vec x_2)\right ]
\label{Eq:SM-entangled-pattern}
\end{eqnarray}
In the coordinates $(\vec r,\vec \rho)$ and integrated over the slit width $a$ the pattern becomes,
 \begin{eqnarray}
\left \langle \mathbf  P^{(2)}(\vec \rho_1, \vec \rho_2  )  \right\rangle_\mathrm{\!ent2}
\! =   \cos^2 \! \left[\frac{k\, \ell}{ 2 z_0} |\vec \rho_1 \! +\! \vec \rho_2|  \right ] \mathrm{sinc}^2 \!\left[  \frac{k\, a}{2 z_0} |\vec \rho_1 \! + \! \vec \rho_2|\right ]
\hspace*{-15pt} \nonumber \\
\label{Eq:SM-entangled-state-pattern}
\end{eqnarray}

\subsection{$N$-Photon Number State}

For the \emph{2-mode $N$-photon number state} $|\phi_\mathrm{numN}\rangle$  the matrix elements of $\mathbf A$ are given by
 \begin{eqnarray}
 \langle \phi_\mathrm{numN}  |\mathbf a^\dag_{\vec{k} } \, \mathbf   a^\dag_{ \vec{k}'} \mathbf a_{\vec{k}' } \, \mathbf   a_{ \vec{k}} |\phi_\mathrm{numN}\rangle = \frac{N^2}{4}
\end{eqnarray}
and for the first two terms in $\mathbf B$ we have,
\begin{eqnarray}
&\,& \langle \phi_\mathrm{numN} |\mathbf a^\dag_{\vec{k} } \, \mathbf   a^\dag_{ \vec{k} } \mathbf a_{\vec{k}  } \, \mathbf   a_{ \vec{k}} |\phi_\mathrm{numN}\rangle
\nonumber \\
&\,&= \langle \phi_\mathrm{numN} |\mathbf a^\dag_{\vec{k}' } \, \mathbf   a^\dag_{ \vec{k}' } \mathbf a_{\vec{k} ' } \, \mathbf   a_{ \vec{k}'} |\phi_\mathrm{numN}\rangle
 \nonumber \\
 &\,& =\frac{N(N-2)}{4}
\end{eqnarray}
Those for the second two terms in $\mathbf B$  and those of  $\mathbf C$ and  $\mathbf D$  vanish. We obtain in general,
\begin{eqnarray}
&\,& \hspace*{-25pt} \left \langle \mathbf  P^{(2)}(\vec x_1, \vec x_2  )  \right\rangle_\mathrm{\!numN}\!=\!\frac{1}{4}\left \langle  \phi_\mathrm{num2} |\mathbf A +\mathbf B|  \phi_\mathrm{num2} \right \rangle
\nonumber \\
&\,&    =  \frac{N }{8}\left\{2N \cos^2 \!\left[\frac{1}{2} (\vec{k} - \vec{k}')\cdot(\vec{x}_1 - \vec{x}_2) \right] \! +\! (N \! - \! 2) \! \right\}
 \hspace*{-15pt}
 \nonumber\\
\end{eqnarray}
For the case $N=2$ double-slit case we obtain,
 \begin{eqnarray}
\hspace*{-15pt} \left \langle \mathbf  P^{(2)}(\vec \rho_1, \vec \rho_2  )  \right\rangle_\mathrm{\!num2}
\!&=&\!   \cos^2 \! \left[\frac{k\, \ell}{ 2 z_0} |\vec \rho_1 \! -\! \vec \rho_2|  \right ]
\nonumber \\
&\,& \times \mathrm{sinc}^2 \!\left[  \frac{k\, a}{2 z_0} |\vec \rho_1 \! - \! \vec \rho_2|\right ]
\label{Eq:SM-number-state-pattern}
\end{eqnarray}

\section{Evaluation of the Degree of Second Order Coherence}

For the \emph{collective} coherent state $|\Phi_\mathrm{coh}\rangle$, the numerator in (\ref{Eq:Second-order-degree-of-coherence}) factors into the denominator so that
\begin{eqnarray}
g^{(2)}_\mathrm{coh}(\vec \rho,-\vec \rho)= 1
\label{Eq:g2-coh-collective}
\end{eqnarray}
For the coherent \emph{substates}  $|\phi_\mathrm{cohN}\rangle$,  we obtain from (\ref{Eq:SM-1-photon-double-slit-cohN}) and (\ref{Eq:SM-2-photon-double-slit-cohN}),
\begin{eqnarray}
g^{(2)}_\mathrm{cohN}(\vec \rho,-\vec \rho)=  1-\frac{1}{N}
\label{Eq:g2-coh-N-substate}
\end{eqnarray}

For the 2-photon entangled state $|\phi_\mathrm{ent2}\rangle$  given by (\ref{Eq:SM-2-mode-ent-state}), we obtain from (\ref{Eq:SM-1P-ent-pattern}) and (\ref{Eq:SM-entangled-state-pattern}),
\begin{eqnarray}
g^{(2)}_\mathrm{ent2}(\vec \rho,-\vec \rho)= 1
\label{Eq:g2-ent-state}
\end{eqnarray}

For the 2-photon number state $|\phi_\mathrm{num2}\rangle$  given by (\ref{Eq:SM-2-mode-num-state}),  we obtain from (\ref{Eq:SM-number-state-pattern}) and $\left \langle   \mathbf  P^{(1)}(\vec \rho , \vec \rho   )  \right\rangle_\mathrm{\!num2}=\left \langle   \mathbf  P^{(1)}(\vec \rho , \vec \rho   )  \right\rangle_\mathrm{\!ent2}$,
\begin{eqnarray}
g^{(2)}_\mathrm{num2}(\vec \rho,-\vec \rho)  =  \cos^2 \! \left[\frac{k\, \ell}{  z_0} \rho \right ] \mathrm{sinc}^2 \!\left[  \frac{k\, a}{z_0}\rho \right ]
\label{Eq:g2-num-state}
\end{eqnarray}

For the collective phase-diffused coherent state $|\Phi_\mathrm{dif}\rangle$, we obtain
\begin{eqnarray}
g^{(2)}_\mathrm{dif}(\vec \rho,-\vec \rho) = \frac{1}{2} + \cos^2 \! \left[\frac{k\, \ell}{  z_0} \rho \right ] \mathrm{sinc}^2 \!\left[  \frac{k\, a}{z_0}\rho \right ]
\label{Eq:g2-dif-state}
\end{eqnarray}
and for the substates $|\phi_\mathrm{difN}\rangle$ we have,
\begin{eqnarray}
g^{(2)}_\mathrm{difN}(\vec \rho,-\vec \rho) = \! \left[1 \!-\!\frac{1}{N}\right] \! \left\{ \frac{1}{2}\! + \!\cos^2 \! \left[\frac{k\, \ell}{  z_0} \rho \right ] \mathrm{sinc}^2 \!\left[  \frac{k\, a}{z_0}\rho \right] \!\right\}
\label{Eq:g2-difN-state}
\nonumber \\
\hspace*{-15pt}
\end{eqnarray}

For the collective chaotic state $|\Phi_\mathrm{cha}\rangle$, we obtain
\begin{eqnarray}
g^{(2)}_\mathrm{cha}(\vec \rho,-\vec \rho) = 1 + \cos^2 \! \left[\frac{k\, \ell}{  z_0} \rho \right ] \mathrm{sinc}^2 \!\left[  \frac{k\, a}{z_0}\rho \right ]
\label{Eq:g2-cha-state}
\end{eqnarray}
and for  the substates $|\phi_\mathrm{chaN}\rangle$ we have
\begin{eqnarray}
g^{(2)}_\mathrm{chaN}(\vec \rho,-\vec \rho) = \! \frac{2}{3}\left[1\!-\!\frac{1}{N}\right]\!\left\{1 \!+\! \cos^2 \! \left[\frac{k\, \ell}{  z_0} \rho \right ] \mathrm{sinc}^2 \!\left[  \frac{k\, a}{z_0}\rho \right ] \!\right\}
\nonumber \\
\hspace*{-15pt}
\label{Eq:g2-chaN-state}
\end{eqnarray}


\begin{thebibliography}{10}
\bibitem{cloning-comment}
In Refs.\,\cite{stohr:17,stohr-AOP} it was assumed that the 2-photon coherent
  or cloned-biphoton state is second order coherent, which the present analysis
  reveals to be incorrect. Second order coherence only develops when the
  stimulation process involves many photons.
\end{thebibliography}
\end{document}